\documentclass[12pt]{article}

\usepackage{epsf,amsfonts,hyperref}
\usepackage{cite}
\renewcommand{\appendix}[1]{
    \addtocounter{section}{1}
    \setcounter{equation}{0}
    \renewcommand{\thesection}{\Alph{section}}
    \section*{Appendix \thesection\protect\indent #1}
    \addcontentsline{toc}{section}{Appendix \thesection\ \ \ #1}
}

\newcommand{\ee}[1]{{\rm e}^{#1}}
\newcommand{\beq}{\begin{equation}}
\newcommand{\eeq}{\end{equation}}
\newcommand{\bea}{\begin{eqnarray}}
\newcommand{\eea}{\end{eqnarray}}

\newcommand{\N}{{\cal N}}

\newcommand{\Tr}{{\,\rm Tr}\:}
\newcommand{\tr}{{\,\rm tr}\:}
\renewcommand{\d}{{{\partial}}}

\newcommand{\refeq}[1]{ eq.(\ref{#1})}
\newcommand{\ul}[1]{ \underline{#1}}
\newcommand{\virg}{\,\, , \quad}

\newcommand{\td}{\tilde}
\newcommand{\CC}{{\mathbf C}}
\newcommand{\Res}{\mathop{{\rm Res}\,}}
\newcommand{\Pol}{{{\rm Pol}\,}}
\newcommand{\Frac}{{{\rm Frac}\,}}
\newcommand{\ds}{\displaystyle}
\newcommand{\Remark}{\medskip {\noindent \bf Remark: }}

\newcommand\encadremath[1]{\vbox{\hrule\hbox{\vrule\kern8pt
\vbox{\kern8pt \hbox{$\displaystyle #1$}\kern8pt}
\kern8pt\vrule}\hrule}}
\def\enca#1{\vbox{\hrule\hbox{
\vrule\kern8pt\vbox{\kern8pt \hbox{$\displaystyle #1$}
\kern8pt} \kern8pt\vrule}\hrule}}

\begin{document}

\begin{flushright}

\hfill Saclay-T03/125

\end{flushright}

\vspace{0.2cm}

\begin{center}

\begin{Large}

\textbf{Master loop equations, free energy and correlations
 for the chain of matrices\footnote{Work partialy supported by the EC IHP Network:
``Discrete geometries: from solid state physics to quantum gravity'', 
HPRN-CT-1999-000161.}}

\end{Large}

\vspace{1.0cm}

\begin{large}


\end{large}

\bigskip

\begin{small}

$^{\star}$ {\em Service de Physique Th\'eorique, CEA/Saclay \\ Orme des

Merisiers F-91191 Gif-sur-Yvette Cedex, FRANCE } \\

\end{small}

\bigskip

\bigskip


{\bf Abstract}

\end{center}

\begin{center}

\begin{small}

\parbox{13cm}{
The loop equations for a chain of hermitian random matrices are computed
explicitely, including the $1/N^2$ corrections.
To leading order, the master loop equation reduces to an algebraic equation,
whose solution can be written in terms of geometric properties of the
underlying algebraic curve.
In particular we compute the free energy, the resolvents, the 2-loop functions
and some mixed one loop functions.
We also initiate the calculation of the $1/N^2$ expansion.
}

\end{small}

\end{center}



%

%

%


%

%

%

%

%


%

\section{Introduction}

The multimatrix model, or chain of matrices is defined by the
probability weight for $\N+1$ hermitean matrices of size $N\times N$:
\beq\label{probaweight}
d\mu(M_0,\dots,M_\N) = {1\over Z}  d M_0 \dots d M_\N \,\,
\ee{-{N\over T}\tr [ \sum_{k=0}^\N V_k(M_k) - \sum_{k=1}^{\N} M_{k-1}M_k ]}
\eeq
\beq\label{defZF}
Z := (N/2\pi)^{-{1\over 2}(\N+1)N^2}\,\,\ee{-{N^2\over T^2} F} 
:= \int 
d M_0 \dots d M_\N \,\,
\ee{-{N\over T}\tr [ \sum_{k=0}^\N V_k(M_k) - \sum_{k=1}^{\N} M_{k-1}M_k ]}
\eeq
where $M_k$ ($k=0,\dots,\N$) are $N\times N$ hermitean matrices, and
$d M_k$ is the product of Lebesgue measures of all real components of $M_k$.
$Z$ is the partition function, $F$ is the free energy.
$T$ is the temperature, it can be chosen equal to $1$ in most of the paper,
except when one is interested in derivatives with respect to $T$.

The potentials $V_k$ are polynomials of degree $d_k+1$:
\beq
V_k(x) = g_{k,0} + \sum_{j=1}^{d_k+1} {g_{k,j}\over j} x^j
\eeq
In order for $Z$ to exist, we have to assume that all $V_k$'s are bounded
from below on the real axis.
However, that constraint can be relaxed by studying "normal matrices"
instead of hermitean matrices\footnote{i.e. the eigenvalues are located on
complex paths instead of the real axis.
For a potential of degree $d_k+1$, there are $d_k$ homologicaly non equivalent
complex paths on which $V_k$ is bounded from below \cite{Marcopaths}.},
or by considering $Z$ as a formal power series in the coefficients of
the potentials.
For simplicity, we assume here that the $V_k$'s are real\footnote{Most of the
results presented here remain valid for complex $V_k's$.
The asumption of real potential merely allows to have a more concise derivation
 of loop equations.}.

Each $M_k$ lies in the potential well $V_k$ and is linearly coupled to its
neighboors $M_{k-1}$ and $M_{k+1}$.
$M_0$ and $M_\N$ have only one neighboor, and thus the chain is open.
So far, the problem of the closed chain has remained unsolved.

The multimatrix model is a generalization of the 2-matrix model.
It was often considered in the context of 2-dimensional quantum gravity
and string theory.
Its critical points are known to represent the minimal conformal
field theories,
characterized by a pair of integers $(p,q)$.
It is known that one can get a $(p,q)$ critical point with a multimatrix model
where $\N=q-2$.
The necessity of studying multimatrix models can be understood from the fact
that the one matrix models contains only the critical models with $q\leq 2$.

Recent progress have been made in the understanding of the two-matrix model
\cite{Bertolafreeenergy, BEHRH, McLaughlin, eynm2m, guionnet, KazMar},
and it has been often noticed that the chain of matrices presents many
similarities with the 2-matrix model \cite{DKK}.

The loop equations of the 2-matrix model have been known for a long time
\cite{eynard, staudacher} and have been concisely written in \cite{eynm2m}
including the $O(1/N^2)$ terms.
The loop equations for the chain of matrices have been written in an appendix of
\cite{eynardchain} and in a draft of \cite{eynardthese} without the $O(1/N^2)$ terms, and without any proof,
and also with misprints.
The present paper is mainly a translation from french to english of
the draft \cite{eynardthese}, with proofs, and additional results.
It thus fills a gap in the appendix of \cite{eynardchain}.
In particular, the $1/N^2$ terms are included, opening the way to finding
next to leading order corrections as in \cite{eynm2m}.
The leading order solution is also presented in an an updated algebraic
geometry language.

\bigskip

The paper is organized as follows:
\begin{enumerate}
\item introduction

\item definitions of the loop functions.

\item the main result and its derivation: the master loop equation.
Readers less interested in technical details can skip the derivation.

\item to leading order, the master loop equation is an algebraic equation.
We discuss its solution.

\item computation of the free energy in the large N limit,
along the lines drawn by \cite{Bertolafreeenergy}.

\item computation of other resolvents and correlators in the large N limit.

\item next to leading order, along the lines drawn by \cite{eynm2m, eynm2mg1}.

\item examples (2 and 3-matrix models, gaussian case, one-cut case).
\end{enumerate}

The most important results of this paper are the master loop equation
\refeq{masterloopeq}, the free energy \refeq{leadingF}, 
the two-loop function \refeq{twoloopbergmann},
the mixed correlators \refeq{mixednextcorrelators}, 
\refeq{W0N} and \refeq{UlargeN}.

\section{Definitions}

\subsection{Moments}

We define the moments:
\beq
T_{n_0,n_1,\dots,n_\N}  :=
{1\over 2N}\left< \tr M_0^{n_0}M_1^{n_1}\dots M_\N^{n_\N} \right>
+ {1\over 2N}\left< \tr M_\N^{n_\N}\dots M_1^{n_1} M_0^{n_0}\right>
\eeq
where $<>$ means averaging with the probability measure of \refeq{probaweight}.
If all the potentials $V_k$'s are bounded from below,
it is clear that the moments are well defined convergent integrals.
If on the contrary the $V_k$'s are not all bounded from below, then the moments
are only formaly defined through their expansions in the coupling constants.

\subsection{The complete one loop-function}

We define the complete one loop-function as follows:
\beq\label{defcompleteoneloopf}
W(z_0,\dots,z_{\N}) := \sum_{n_0,\dots,n_\N=0}^\infty
{T_{n_0,n_1,\dots,n_\N}\over z_0^{n_0+1} z_1^{n_1+1} \dots z_\N^{n_\N+1}}
\eeq
which can also be written
\bea
W(z_0,\dots,z_{\N}) & = & {1\over 2N}
\left< \tr {1\over z_0-M_0}{1\over z_1-M_1}\dots{1\over z_\N-M_\N} \right> \cr
&& +{1\over 2N}
\left< \tr {1\over z_\N-M_\N}\dots{1\over z_1-M_1}{1\over z_0-M_0} \right>
\eea
Notice that it is only a formal definition (even when all the potentials
are bounded from below),
it makes sense only through its power expansion in the vicinity of
all $z_k\to\infty$.
That function is merely a convenient concise notation for dealing with all
the moments at once.

\subsection{Uncomplete one loop-functions}

We also define ``uncomplete'' loop-functions.
For any integer $k$ between zero and $\N+1$, and for any ordered subset
of $[0,\N]$
$I=\{i_1,\dots,i_k\}$ ($0\leq i_1<i_2<\dots<i_k\leq \N$), we define:

\bea\label{defuncompleteoneloopf}
W_{i_1,\dots,i_k}(z_{i_1},\dots,z_{i_k})
 := & {1\over 2N}\left< \tr {1\over z_{i_1}-M_{i_1}}
 {1\over z_{i_2}-M_{i_2}}\dots{1\over z_{i_k}-M_{i_k}} \right> \cr
& +{1\over 2N}\left< \tr {1\over z_{i_k}-M_{i_k}}\dots
{1\over z_{i_2}-M_{i_2}}{1\over z_{i_1}-M_{i_1}} \right> \cr
 = & (-1)^{\N+1-k} \ds\Res \prod_{j\notin I}\, dz_j\,  W(z_0,\dots,z_\N)  \cr
 = & \ds\mathop\Pol_{z_j,\,\, j\notin I} W(z_0,\dots,z_\N)\,
 \prod_{j\notin I} z_j
\eea
where $\Pol$ means the polynomial part in the vicinity of $\infty$.

\subsection{The resolvents}

As a special case of \refeq{defuncompleteoneloopf},
the resolvent of the $k^{\rm th}$ matrix of the chain is defined by:
\beq\label{defWk}
W_k(z_k) := {1\over N}\left< \tr {1\over z_k-M_{k}} \right>
\eeq
In particular, the resolvent of the first matrix of the chain will play a
prominent role:
\beq\label{defW0}
W_0(z_0) := {1\over N}\left< \tr {1\over z_0-M_{0}} \right>
\eeq
The master loop equation derived in section \ref{masterloopeqsection}
is an equation for $W_0(z_0)$ as a function of $z_0$,
and in section \ref{largeNsection},
we determine the large $N$ limit of $W_0(z_0)$.
In section \ref{otherresolvents},
we derive the large $N$ limits of the other resolvents $W_k$.

Instead of $W_0(z_0)$, it appears more convenient to consider
the following function:
\beq
Z_1(z_0) := V'_0(z_0) - T W_0(z_0)
\eeq
and more generaly we define:
\beq\label{defZk}
\displaystyle
\left\{
\begin{array}{l}
Z_{-1}(z_0) := T W_0(z_0)
\cr
Z_0(z_0):=z_0
 \cr
Z_{k+1}(z_0) := V'_k(Z_k(z_0)) - Z_{k-1}(z_0) \qquad 1\leq k\leq \N
\end{array}
\right.
\eeq
For short, we write:
\beq
Z_k := Z_k(z_0)
\eeq
Notice that all $Z_k$'s are polynomials in both $z_0$ and $W_0(z_0)$,
and therefore, any polynomial of the $Z_k$'s is a polynomial in $z_0$
and $W_0(z_0)$.

\subsection{Polynomial loop functions $P$ and $U$}

We define the functions $f_{k,l}(z_k,\dots,z_l)$:
\beq\label{deffkl}
\displaystyle
\left\{
\begin{array}{l}
f_{k,l}  :=  0 \qquad {\rm if}\,\,  l<k-1 \cr
f_{k,k-1}  :=  1  \cr
f_{k,k}(z_k)  :=  V'_k(z_k) \cr
f_{k,l+1}(z_k,\dots,z_{l+1})  :=  V'_{l+1}(z_{l+1}) f_{k,l}(z_k,\dots,z_l)
- z_l z_{l+1} f_{k,l-1}(z_k,\dots,z_{l-1})
\end{array}
\right.
\eeq
Then we define for $k\leq l$:
\beq\label{defPkl}
P_{k,l}(z_0,\dots,z_\N) := \mathop\Pol_{z_k,\dots,z_l} f_{k,l}(z_k,\dots,z_l)
W(z_0,\dots,z_\N)
\eeq
where $\Pol$ means the polynomial part in the vicinity of $\infty$.
$P_{k,l}$ depends on $\N+1$ variables, and is polynomial in the variables
$z_k,\dots,z_l$.

By definition of $f_{k,l}$ we have:
\beq\label{recurPkl}
P_{k,l}(z_0,\dots,z_\N)
=
\mathop\Pol_{z_l} V'_l(z_l) P_{k,l-1}(z_0,\dots,z_\N)
- \mathop\Pol_{z_l,z_{l-1}} z_{l-1}\,z_l\,\, P_{k,l-2}(z_0,\dots,z_\N)
\eeq

In particular, we define:
\beq\label{defP}
P(z_0,\dots,z_\N) := P_{0,\N}(z_0,\dots,z_\N)
\eeq
which is polynomial in all the variables,
and
\beq\label{defU}
U(z_0,\dots,z_\N) := P_{1,\N}(z_0,\dots,z_\N)
\eeq
which is polynomial in all the variables but $z_0$.

We also define:
\beq\label{defE}
E(z_0,\dots,z_\N) := (V'_0(z_0)-z_1)(V'_\N(z_\N)-z_{\N-1})
- T P(z_0,\dots,z_\N)
\eeq
which is also a polynomial in all variables.
These polynomials will play an important role in the master loop equation.

\subsection{2-loop functions}

We define the loop insertion operator as in \cite{ACKM}:
\beq\label{defloopinsert}
{\d \over \d V_l(z)} := -{1\over z}{\d\over \d g_{l,0}}-\sum_{k=1}^\infty
{k\over z^{k+1}} {\d\over \d g_{l,k}}
\eeq
where $V_l(z)=g_{l,0}+\sum_{k\geq 1} {g_{l,k}\over k} z^{k}$,
and the derivatives are
taken at $g_{l,k}=0$ for $k>d_l+1$.
It is a formal definition, which makes sense only order by order in
its large $z$ expansion.

It is such that (for $|z|>|z'|$):
\beq\label{dVdV}
{\d V_k(z')\over \d V_l(z)} = -\delta_{k,l} {1\over z-z'}
\virg
{\d V'_k(z')\over \d V_l(z)} = -\delta_{k,l} {1\over (z-z')^2}
\eeq
When applied to \refeq{defZF}, it produces a resolvent,
i.e. the expectation of a trace:
\beq\label{dFdV}
{\d F\over \d V_l(z)} = -T W_l(z) =
-{T\over N}\left< \tr {1\over z-M_{l}} \right>
\eeq
and more generaly, the action of ${\d \over \d V_l(z)}$ on an expectation
value inserts a new trace.

In particular, we define the following two loop functions (i.e. two-traces):
\bea\label{defW2loop}
W_{;l}(z_0,\dots,z_{\N};z) &:=& T{\d \over \d V_l(z)} W(z_0,\dots,z_{\N}) \cr
& = & {1\over 2}\left<
\tr {1\over z_0-M_0}{1\over z_1-M_1}\dots{1\over z_\N-M_\N}
\tr {1\over z-M_l}\right> \cr
&& +{1\over 2}\left< \tr {1\over z-M_l}
\tr {1\over z_\N-M_\N}\dots{1\over z_1-M_1}{1\over z_0-M_0} \right> \cr
&& - N^2 W_l(z)W(z_0,\dots,z_{\N})
\eea
\bea\label{defWI2loop}
W_{i_1,\dots,i_k;l}(z_{i_1},\dots,z_{i_k};z)
& := & T{\d \over \d V_l(z)} W_{i_1,\dots,i_k}(z_{i_1},\dots,z_{i_k}) \cr
& = & (-1)^{\N+1-k}\ds\Res \prod_{j\notin I}\, dz_j\,  W_{;l}(z_0,\dots,z_\N;z)
\eea
\beq\label{defPkl2loop}
P_{k,l;j}(z_0,\dots,z_\N;z)
:= \mathop\Pol_{z_k,\dots,z_l} f_{k,l}(z_k,\dots,z_l) W_{;j}(z_0,\dots,z_\N;z)
\eeq
\beq\label{defP2loop}
P_{;j}(z_0,\dots,z_\N;z) := P_{0,\N;j}(z_0,\dots,z_\N;z)
\eeq
\beq\label{defU2loop}
U_{;j}(z_0,\dots,z_\N;z) := P_{1,\N;j}(z_0,\dots,z_\N;z)
\eeq
In particular, the following two-loop functions will play an important role:
\bea\label{defWkl2loop}
W_{k;l}(z';z) & = &T{\d W_k(z')\over \d V_l(z)}
=-{\d^2 F\over \d V_k(z')\d V_l(z)}
 =  T{\d W_l(z)\over \d V_k(z')} = W_{l;k}(z;z') \cr
& = & \left< \tr {1\over z-M_l}\tr {1\over z'-M_k} \right>_{\rm c}
\eea
where the subscript c means connected part $<AB>_{\rm c} := <AB>-<A><B>$.

\section{Loop equations}

\subsection{Main result: the master loop equation}
\label{masterloopeqsection}

We prove below that the master loop equation
(named after \cite{staudacher}) can be written:
\beq\label{masterloopeq}
\encadremath{
E(Z_0,Z_1,\dots,Z_\N)
= {T^2\over N^2} U_{;0}(Z_0,Z_1,\dots,Z_\N;Z_0)
}\eeq
where the functions $Z_k(z_0)$ have been defined in \refeq{defZk}.
Examples for $\N=1$ (the 2-matrix model) and $\N=2$ (the 3-matrix model)
are explicited in section \ref{examples}.

As an intermediate step in proving \refeq{masterloopeq},
we need to prove the following formula for all $1\leq k\leq \N$:
\beq\label{Wkrechypothesis}
\encadremath{
\begin{array}{ll}
 (z_k-Z_k)W_{0,k,\dots,\N}(z_0,z_k,\dots,z_\N)
= &  W_{0,k+1,\dots,\N}(z_0,z_{k+1},\dots,z_\N) \cr
& - P_{0,k-1}(z_0,Z_1,\dots,Z_{k-1},z_k,\dots,z_\N) \cr
& -{T\over N^2} P_{1,k-1;0}(z_0,Z_1,\dots,Z_{k-1},z_k,\dots,z_\N;z_0) \cr
\end{array}
}
\eeq

\subsection{Derivation of the master loop equation}

\begin{itemize}

\item Proof of \refeq{Wkrechypothesis} for $k=1$:

The invariance of the matrix integral under the infinitesimal change of
variable (see \cite{courseynard, ZJDFG, eynm2m}, and pay attention to
the non-commutativity of matrices,
and to the hermiticity of the change of variables):
\bea
\delta M_0 & = & {T\over 2} {1\over z_1-M_1}{1\over z_2-M_2}
\dots{1\over z_\N-M_\N}
{1\over z_0-M_0}\cr
&&+ {T\over 2} {1\over z_0-M_0}{1\over z_\N-M_\N}\dots{1\over z_2-M_2}
{1\over z_1-M_1}
\eea
implies
\bea
&& TW_0(z_0)W(z_0,z_1,\dots,z_\N)
+{T\over N^2}W_{;0}(z_0,z_1,\dots,z_\N;z_0) \cr
&& =
V'_0(z_0)W(z_0,z_1,\dots,z_\N)
- P_{0,0}(z_0,z_1,\dots,z_\N) \cr
&& -z_1 W_{0,1,\dots,\N}(z_0,z_1,\dots,z_\N)
+ W_{0,2,\dots,\N}(z_0,z_2,\dots,z_\N)\cr
\eea
The LHS comes from the Jacobian of the change of variable,
and the RHS comes from the variation of the action.
The first two terms of the RHS come from $\delta \tr V_0(M_0)
= \tr V'_0(M_0)\delta M_0$,
and use of \refeq{formulaVzM}.
The last two terms of the RHS come from $\delta \tr M_0 M_1
= \tr M_1\delta M_0$,
and use of \refeq{formulaMzM}.

Using $Z_1=V'_0(z_0)-TW_0(z_0)$, this can be rewritten:
\bea\label{loopeqWzero}
(z_1-Z_1)W_{0,1,\dots,\N}(z_0,z_1,\dots,z_\N)
& = & W_{0,2,\dots,\N}(z_0,z_2,\dots,z_\N) \cr
& & - P_{0,0}(z_0,z_1,\dots,z_\N) \cr
&& -{T\over N^2}W_{;0}(z_0,z_1,\dots,z_\N;z_0)
\eea
Therefore we have derived \refeq{Wkrechypothesis} in the case $k=1$.

\item Proof of \refeq{Wkrechypothesis} for $k=2$:

Notice that \refeq{loopeqWzero} implies:
\bea\label{loopeqVun}
-\Res dz_1\, W(z_0,z_1,z_2,\dots,z_\N) V'_1(z_1)
& = & V'_1(Z_1) W_{0,2,\dots,\N}(z_0,z_2,\dots,z_\N) \cr
&& - P_{0,1}(z_0,Z_1,z_2,\dots,z_\N) \cr
&& - W_{2,\dots,\N}(z_2,\dots,z_\N) \cr
&& -{T\over N^2} P_{1,1;0}(z_0,Z_1,z_2,\dots,z_\N;z_0) \cr
\eea

The change of variable:
\beq
\delta M_1  =  {1\over 2} {1\over z_2-M_2}\dots{1\over z_\N-M_\N}
{1\over z_0-M_0}
+ {1\over 2} {1\over z_0-M_0}{1\over z_\N-M_\N}\dots
{1\over z_2-M_2}
\eeq
gives (since $\delta M_1$ is independant of $M_1$ there is no jacobian,
the LHS is zero, and the RHS is the variation of
$\tr V'_1(M_1)-M_0M_1 - M_1 M_2$):
\bea
0 & = & -\Res dz_1\, W(z_0,z_1,z_2,\dots,z_\N) V'_1(z_1) \cr
&& - z_0 W_{0,2,\dots,\N}(z_0,z_2,\dots,z_\N)
+ W_{2,\dots,\N}(z_2,\dots,z_\N) \cr
&& - z_2 W_{0,2,\dots,\N}(z_0,z_2,\dots,z_\N)
+ W_{0,3,\dots,\N}(z_0,z_3,\dots,z_\N) \cr
\eea
i.e., using \refeq{loopeqVun} and $V'_1(Z_1)=z_0+Z_2$:
\bea
(z_2-Z_2) W_{0,2,\dots,\N}(z_0,z_2,\dots,z_\N)
& = & W_{0,3,\dots,\N}(z_0,z_3,\dots,z_\N) \cr
&& - P_{0,1}(z_0,Z_1,z_2,\dots,z_\N) \cr
&& -{T\over N^2} P_{1,1;0}(z_0,Z_1,z_2,\dots,z_\N;z_0) \cr
\eea
Therefore we have derived \refeq{Wkrechypothesis} in the case $k=2$.

\item Proof of \refeq{Wkrechypothesis} by recursion on $k$:

We have already proved \refeq{Wkrechypothesis} for $k=1$ and $k=2$.
Now assume that \refeq{Wkrechypothesis} holds for $k-1$ and $k$,
we are going to prove it for $k+1$.

Note that \refeq{Wkrechypothesis} for $k-1$ implies
(multiply by $z_{k-1}$ and take the residue at $z_{k-1}\to\infty$
and $z_k\to\infty$):
\bea\label{loopeqdotk}
&&
\Res dz_{k-1} \Res dz_k \,\, z_{k-1}
W_{0,k-1,k,\dots,\N}(z_0,z_{k-1},z_{k},z_{k+1},\dots,z_\N) \cr
 = && Z_{k-1} W_{0,k+1,\dots,\N}(z_0,z_{k+1},\dots,z_\N) \cr
&& - \mathop\Pol_{z_{k-1},z_k} \left.
z_{k-1}z_k P_{0,k-2}(z_0,\dots,z_\N)
\right|_{z_0=Z_0,\dots,z_{k}=Z_{k}} \cr
&& -{T\over N^2} \mathop\Pol_{z_{k-1},z_{k}} \left.
z_k z_{k-1} P_{1,k-2;0}(z_0,\dots,z_\N;z_0)
\right|_{z_0=Z_0,\dots,z_{k}=Z_{k}}\cr
\eea
and \refeq{Wkrechypothesis} for $k$ implies (multiply by $V'_k(z_k)$ and
take the residue at $z_k\to\infty$):
\bea\label{loopeqVk}
&& -\Res dz_k \,\, V'_k(z_k)\,
W_{0,k,k+1,\dots,\N}(z_0,z_k,z_{k+1},\dots,z_\N) \cr
= && V'_k(Z_k) W_{0,k+1,\dots,\N}(z_0,z_{k+1},\dots,z_\N) \cr
&& - \mathop\Pol_{z_k} \left.
V'_k(z_k)P_{0,k-1}(z_0,\dots,z_\N)
\right|_{z_0=Z_0,\dots,z_{k}=Z_{k}} \cr
&& -{T\over N^2} \mathop\Pol_{z_{k}} \left.
V'_k(z_k) P_{1,k-1;0}(z_0,\dots,z_\N;z_0)
\right|_{z_0=Z_0,\dots,z_{k}=Z_{k}}\cr
\eea

Then consider the change of variable:
\bea
\delta M_k  & = &  {1\over 2} {1\over z_{k+1}-M_{k+1}}\dots{1\over z_\N-M_\N}
{1\over z_0-M_0} \cr
&& + {1\over 2} {1\over z_0-M_0}{1\over z_\N-M_\N}\dots
{1\over z_{k+1}-M_{k+1}}
\eea
it gives
\bea
0 & = &
-\Res dz_k \,\, V'_k(z_k)\,
W_{0,k,k+1,\dots,\N}(z_0,z_k,z_{k+1},\dots,z_\N) \cr
&& - \Res dz_{k-1} \Res dz_k \,\, z_{k-1}
W_{0,k-1,\dots,\N}(z_0,z_{k-1},\dots,z_\N)  \cr
&& - z_{k+1} W_{0,k+1,\dots,\N}(z_0,z_{k+1},\dots,z_\N)
+ W_{0,k+2,\dots,\N}(z_0,z_{k+2},\dots,z_\N) \cr
\eea
using \refeq{loopeqdotk} and \refeq{loopeqVk},
as well as $V'_k(Z_k) = Z_{k-1}+Z_{k+1}$,
we get \refeq{Wkrechypothesis} for $k+1$.

\item \refeq{Wkrechypothesis} for $k=\N$:

In particular, the previous recurrence derivation shows that
\refeq{Wkrechypothesis} for $k=\N$ reads:
\beq\label{WkrechypothesisN}
\begin{array}{ll}
 (z_\N-Z_\N)W_{0,\N}(z_0,z_\N)
= &  W_{0}(z_0) \cr
& - P_{0,\N-1}(z_0,Z_1,\dots,Z_{\N-1},z_\N) \cr
& -{T\over N^2} P_{1,\N-1;0}(z_0,Z_1,\dots,Z_{\N-1},z_\N;z_0) \cr
\end{array}
\eeq

\item Proof of \refeq{masterloopeq}:

In particular, from \refeq{Wkrechypothesis} for $k=\N-1$ we derive:
\bea\label{loopeqdotNmun}
&& \Res dz_{\N-1} \Res dz_{\N} \,\, z_{\N-1}
W_{0,z_{\N-1},z_\N}(z_0,z_{\N-1},z_\N) \cr
& = & Z_{\N-1} W_0(z_0) \cr
&& - \mathop\Pol_{z_{\N-1},z_{\N}}
\left. z_{\N-1}z_{\N} P_{0,\N-2}(z_0,\dots,z_\N)
\right|_{z_0=Z_0,\dots,z_{\N}=Z_{\N}}  \cr
&& -{T\over N^2} \mathop\Pol_{z_{\N-1},z_{\N}} \left.
z_{\N-1}z_{\N} P_{1,\N-1;0}(z_0,\dots,z_\N;z_0)
\right|_{z_0=Z_0,\dots,z_{\N}=Z_{\N}}\cr
\eea
and from \refeq{WkrechypothesisN}, we derive:
\bea\label{loopeqVN}
&& -\Res dz_\N \,\, V'_\N(z_\N)\, W_{0,\N}(z_0,z_\N) \cr
& = & V'_\N(Z_\N) W_0(z_0) \cr
&& - \mathop\Pol_{z_{\N}}
\left. V'_\N(z_\N) P_{0,\N-1}(z_0,\dots,z_\N)
\right|_{z_0=Z_0,\dots,z_{\N}=Z_{\N}} \cr
&& -{T\over N^2} \mathop\Pol_{z_{\N}} \left.
V'_\N(z_\N) P_{1,\N-1;0}(z_0,\dots,z_\N;z_0)
\right|_{z_0=Z_0,\dots,z_{\N}=Z_{\N}}\cr
\eea

Then, the change of variables
\beq
\delta M_\N = {1\over z_0-M_0}
\eeq
gives
\bea
0 & = &
-\Res dz_\N \,\, V'_\N(z_\N)\, W_{0,\N}(z_0,z_\N) \cr
& & - \Res dz_{\N-1} \Res dz_{\N} \,\, z_{\N-1}
W_{0,z_{\N-1},z_\N}(z_0,z_{\N-1},z_\N)
\eea
i.e., using \refeq{loopeqdotNmun} and \refeq{loopeqVN}:
\bea
0 & = & (V'_\N(Z_\N)- Z_{\N-1}) W_0(z_0) \cr
&& - P_{0,\N}(Z_0,\dots,Z_\N) \cr
&& -{T\over N^2} P_{1,\N;0}(Z_0,\dots,Z_\N;z_0) \cr
\eea
Recalling that $Z_1=V'_0(z_0)-TW_0(z_0)$, we get:
\bea
&&(V'_0(Z_0)-Z_1)(V'_\N(Z_\N)-Z_{\N-1})
- TP_{0,\N}(Z_0,Z_1,\dots,Z_\N) \cr
&&  = {T^2\over N^2} P_{1,\N;0}(Z_0,Z_1,\dots,Z_\N;Z_0)
\eea
which ends the proof of \refeq{masterloopeq}.

\end{itemize}

\section{Large $N$ leading order, algebraic equation}
\label{largeNsection}

Throughough this section and sections 5 and 6, we abusively denote 
with the same name, the loop functions and their large $N$ limits.

\bigskip

To leading order, the master loop equation reduces to an algebraic equation for $W_0$
as a function of $z_0$:
\beq
E(z_0,Z_1,\dots, Z_\N) =0
\eeq
where $Z_1=V'_0(z_0)-TW_0(z_0)$ and $Z_{k+1}=V'_k(Z_k)-Z_{k-1}$.

$E$ is a polynomial of given degrees in each variable, and with known
leading coefficients.
The problem is that most of the subleading coefficients of
$E(z_0,z_1,\dots,z_{\N})$ are not determined by the loop equations.
They are determined by additional hypothesis, which will be explained
in section \ref{cuthypothesis}.
Prior to that, we need to study the geometry of that algebraic equation.

The complex curve $W_0$ (equivalently $Z_1$)
as a function of $z_0$ is a one--dimensional submanifold of $\CC^{\N+1}$,
which is the intersection of $\N$ dimension $\N$ algebraic submanifolds.
The curve $W_0$ as a function of $z_0$ is thus a Riemann surface ${\cal{E}}$.

Instead of viewing the $Z_k$'s as (mutivalued) functions of $z_0$,
it is more appropriate to view the $Z_k$'s as (monovalued) complex functions 
over ${\cal{E}}$:
\beq
p \to z_k(p)
\eeq
such that for all $p\in {\cal E}$:
\beq
Z_k(z_0(p)) = z_k(p)
\eeq
Since the function $z_0(p)$ is not injective, i.e. the point $p$ such that 
$z_0(p)=z_0$ is not unique,
the $Z_k$'s are multivalued functions of $z_0$.
Similarly, one could consider any $z_k$ as a multivalued function of any $z_l$.

Instead of dealing with multivalued functions, we slice ${\cal E}$ into 
domains called sheets,
such that in each domain the functions we are considering are bijections.
The $z_0$-sheets are thus domains in which the function $z_0(p)$ is injective.
More generaly, the $z_k$-sheets are domains in which the function $z_k(p)$
is injective.

Let us study the $z_0$-sheets first.
For that purpose, we define:
\beq\label{defrksk}
\begin{array}{l}
\left\{\begin{array}{lcl}
r_{-1} & := & -1 \cr
r_{0} & := & 1 \cr
r_k & := & d_0 d_1\dots d_{k-1}\,\, , \quad k=1,\dots,\N+1
\end{array}
\right. \cr
\left\{\begin{array}{lcl}
s_{\N+1} & := & -1 \cr
s_{\N} & := & 1 \cr
s_k & := & d_{k+1}\dots d_{\N-1} d_\N\,\, , \quad k=-1,\dots,\N-1
\end{array}
\right.
\end{array}
\eeq

\subsection{$z_0$-sheets}

The equation
\beq
E(z_0,Z_1,\dots, Z_\N) =0
\eeq
has degree $1+d_1d_2\dots d_\N$ in $W_0$ and degree $d_0d_1\dots d_{\N}$
in $z_0$.
Therefore, $W_0(z_0)$ (or equivalently $Z_1$) is a multivalued function
of $z_0$ which takes $1+d_1d_2\dots d_\N$ values:
in other words, there are $1+d_1d_2\dots d_\N\,$ $z_0$-sheets.
We identify these sheets by looking at the asymptotics of
$Z_\N$ when $z_0\to\infty$.

\begin{itemize}
\item{The physical sheet}

From \refeq{defW0}, there must exist at least one solution of the algebraic
equation such that:
\beq\label{W0asymp}
W_0(z_0) \mathop\sim_{z_0\to\infty} {1\over z_0} + O(1/z_0^2)
\eeq
which implies that:
\beq\label{asympz0physsheet}
z_k = O(z_0^{r_k})
\eeq
and in particular
\beq
z_\N = O(z_0^{d_0 d_1\dots d_{\N-1}})
\eeq
The $z_0$-sheet in which these asymptotics hold is called the $z_0$-physical
sheet.

\item{Other sheets}

Notice that the equation for $W_\N(z_\N)$ as a function of $z_\N$
is the same algebraic equation.
In other words, there is a solution of the algebraic equation which is such
that:
\beq\label{asympz0othersheets}
z_k = O(z_\N^{s_k})
\virg
s_k = d_{k+1}\dots d_{\N-1}d_\N
\eeq
i.e.
\beq
z_\N = O(z_0^{1/d_{1}\dots d_{\N-1}d_\N })
\virg
z_k = O(z_0^{1/d_{1}\dots d_{k} })
\eeq
Since the number of $s_0^{\rm th}$ roots of unity is exactly
$s_0=d_{1}\dots d_{\N-1}d_\N$,
we have all the solutions.

\end{itemize}

We thus have $r_0+s_0=1+d_{1}\dots d_{\N-1}d_\N$ sheets.
In one of them we have $z_\N\sim O(z_0^{r_\N})$,
and in the $d_{1}\dots d_{\N-1}d_\N$ others we have
$z_0\sim O(z_\N^{s_0})$.

\subsection{Algebraic geometry}

We now consider the algebraic curve ${\cal E}$ in a more geometric
language.
An abstract point $p\in{\cal E}$ can be represented as a couple $(z_0,z_1)$
such that $z_1=Z_1(z_0)$,
or by any other parametrization.
For instance it can be described as a point of $\CC^\N$,
at the intersection of $\N$ codimension 1 manifolds.

Algebraic geometry is an active and important part of mathematics,
and lots of tools have been invented to describe the geometry of algebraic
Riemann surfaces.
We refer the reader to \cite{Farkas, Fay} for an introduction.

\medskip

A Riemann surface is locally homomorphic to the complex plane $\CC$,
that means that small domains of ${\cal E}$ can be maped on small
domains of $\CC$.
The map, which is one-to-one and holomorphic in that domain is called
a local parameter, let us
call it:
\beq
p\in{\cal E}\to x(p)\in \CC
\eeq
Any complex valued analytic function on ${\cal E}$ can be localy represented
by an analytic function of $x(p)$.

For instance, $z_0=z_0(p)$ is often a good local parameter on ${\cal E}$,
except when $z_0$ approaches a singularity.
Therefore all functions on ${\cal E}$ can be localy written as functions of
$z_0$.
They can also be locally written as functions of any $z_k$.

Notice that the function $z_0(p)$ is not injective, it takes the same value
$z_0$ for different points $p$, namely the same value of $z_0$ corresponds
to $1+s_0$ points $p$.
Therefore the function $z_1(p)$ which is a well defined monovalued function on
${\cal E}$ is a multivalued function of $z_0$.
The $z_0$-sheets are domains of ${\cal E}$ where the function $z_0(p)$ is
injective.
In particular in each sheet, there is only one point $p_\infty$ where
$z_0\to\infty$.

\medskip
Let $p_{\infty+}$ be the point in the physical sheet,
such that $z_0(p_{\infty+})=\infty$.
In the vicinity of $p_{\infty+}$, we have \refeq{asympz0physsheet}:
\beq
z_k(p)\mathop\sim_{p\to p_{\infty+}} z_0^{r_k}(p)
\eeq
i.e. $Z_k(z_0)$ is analytical, which indicates that $z_0$ is a good
local parameter near $p_{\infty+}$.

Now, let $p_{\infty-}$ be the point at $\infty$ in the $z_\N$-physical sheet,
i.e. where
$z_\N(p_{\infty-})=\infty$ and with behavior \refeq{asympz0othersheets}:
\beq
z_k(p)\mathop\sim_{p\to p_{\infty-}} z_\N^{s_k}(p)
\eeq
i.e. $Z_k(z_\N)$ is analytical, which indicates that $z_\N$ is a good
local parameter near $p_{\infty-}$.
In the vicinity of $p_{\infty-}$, the function $z_0(p)$ takes the same value
$z_0$,
$s_0$ times, therefore there are $s_0$ sheets which meet at $p_{\infty-}$.
It is clear that $z_0$ is not a local parameter near $p_{\infty-}$,
but $z_\N$ is.

It is also clear that there can be no other point $p$ such that
$z_0(p)=\infty$.
The algebraic curve ${\cal{E}}$ has only two points at $\infty$.

Notice that the intermediate $z_k$'s with $0<k<\N$ are not appropriate local
coordinates near $p_{\infty+}$ neither near $p_{\infty-}$ (unless many of
the $d_k$'s are equal to $1$).

\medskip
Let us summarize as follows:\\
- The function $z_k(p)$ has a pole of degree $r_k=d_0 d_1 \dots d_{k-1}$
near $p_{\infty+}$. $z_0(p)$ is a local parameter near $p_{\infty+}$.\\
- The function $z_k(p)$ has a pole of degree $s_k=d_{k+1}\dots d_{\N-1} d_{\N}$
near $p_{\infty-}$. $z_\N(p)$ is a local parameter near $p_{\infty-}$.\\
- The function $z_k(p)$ has no other pole.

\Remark
From \refeq{W0asymp}, it is easy to prove by recursion that:
\beq\label{residues}
\Res_{p_{\infty+}} z_{k-1} dz_k = -T = -\Res_{p_{\infty+}} z_{k+1} dz_k
\virg
\Res_{p_{\infty-}} z_{k+1} dz_k = -T = -\Res_{p_{\infty-}} z_{k-1} dz_k
\eeq

\subsubsection{Genus and cycles}

Let $g$ be the genus of ${\cal{E}}$.
It can be proved (using the Riemann-Roch theorem, or the method of
\cite{KazMar}) that
\beq\label{maxgenus}
g\leq g_{\rm max} :=\prod_{k=0}^\N d_k \,\,\, -1
\eeq
Notice that the polynomial $P$ has $1+g_{\rm max}$ coefficients, and
its leading coefficient is fixed,
i.e. $g_{\rm max}$ is the number of coefficients of $P$ not
fixed by the loop equations.

Let  ${\cal{A}}_i , {\cal{B}}_i (i=1,\dots,g)$ be a canonical basis
of non-trivial cycles on ${\cal{E}}$:
\beq
{\cal{A}}_i \bigcap {\cal{B}}_j = \delta_{ij}
\eeq
the choice of non-trivial cycles is not unique, and we will see below one
possible convenient choice.

\subsubsection{Endpoints}

If $z_l$ is considered as a function of $z_k$ ($k\neq l$),
it has singularities (branch points) everytime that:
\beq
dz_k(p)=0
\eeq
Indeed at such a point we have: $z'_l(z_k)=dz_l/dz_k \to \infty$.

The zeroes of $dz_k(p)$ are called the endpoints, and are noted:
\beq
e_{k,i}\,\, , \quad (i=1,\dots,r_k+s_k +2g)
\eeq

Generically, the zeroes of $dz_k$ are simple zeroes and they are all distinct,
which means that $z_k$ behaves as a quadratic function of a local
parameter $x$,
while $z_l$ behaves linearly in $x$.
Therefore, $z_l(z_k)$ has a square root branch point near $z_k(e_{k,i})$.

\subsubsection{Critical points}

It may happen that some endpoints coincide.
This is called a critical point. It is not a generic situation,
it happens only if the potentials $V_k$'s are fine tuned to some
critical potentials.
The critical points are relevant for finding the representations of $(p,q)$
minimal conformal models.
A $(p,q)$ critical point, is such that there exist $k\neq l$,
and a point $e\in{\cal E}$
such that: $dz_k$ has a zero of degree $p-1$ at $e$ and $dz_l$ has
a zero of degree $q-1$ at $e$.
Near $e$, $z_k$ as a function of $z_l$ behaves with a $p/q$ exponent.

\smallskip

From now on, we assume that the potentials $V_k$ are not critical,
i.e. the endpoints are all simple and distinct.

\subsubsection{Cuts}
\label{sectioncuts}

The cuts in ${\cal E}$ are the contours which border the sheets.
Viewed in the $z_k$-plane they are lines in $\CC$ joining two endpoints.

That choice of lines (i.e. the choice of sheets) is largely arbitrary,
provided that they indeed go through the endpoints.

There is a canonical choice, which comes from \cite{eynardchain},
given by the contours where the asymptotics of the biorthogonal polynomials
and their Fourier and Hilbert transforms,
have discontinuities, i.e. some Stokes lines.
That canonical choice is defined as follows:
the set of $z_k$-cuts (contours on ${\cal E}$) is the set of points
$p\in {\cal E}$
such that there exists $p'\neq p$ with $z_k(p')=z_k(p)$ and
\beq
\Re \left( \int_{p}^{p'} Z_{k+1} d z_k \right)=0
\eeq
remark that for all $p$ and $p'$ such that $z_k(p)=z_k(p')$ we have:
\beq
\int_{p}^{p'} z_{k+1} d z_k = \int_{p}^{p'} (V'_k(z_k)-z_{k-1}) d z_k
= -\int_{p}^{p'} z_{k-1} d z_k
\eeq
so that the canonical cuts are left unchanged if we reverse the order of
the chain.

\subsection{Sheet geometry}

\begin{figure}
{\mbox{\epsfxsize=14.truecm\epsfbox{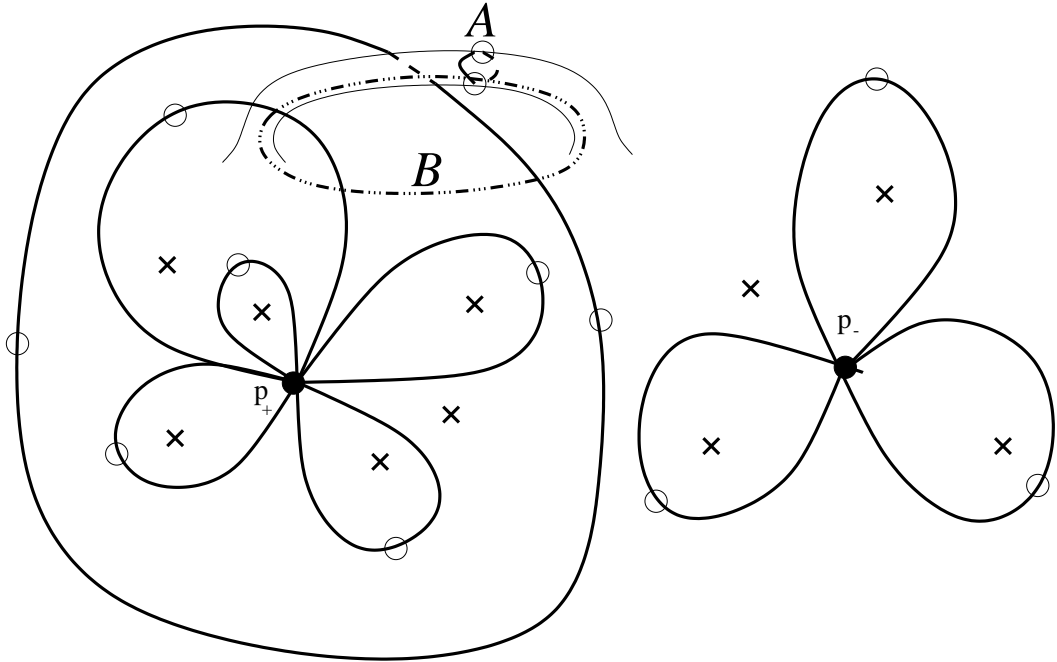}}}
\label{sheetszk} 
\caption{Sheets for $z_k$:
example with $r_k=6$, $s_k=4$ and $g=1$.
The two black cirlces represent $p_{\infty+}$ and  $p_{\infty-}$.
The empty cirles are the $r_k+s_k+2g$ endpoints.
The thick lines, going through the endpoints,
are the cuts separating the sheets.
Each region separated by lines is a sheet, there are $r_k+s_k$ sheets.
The sheets form ``flowers'' near $p_{\infty+}$ and  $p_{\infty-}$ (the petals
have an angle $2\pi/r_{k}$ near $p_{\infty+}$ and $2\pi/s_k$ near
$p_{\infty-}$).
There are $g$ ``handles'', with ${\cal A}$ and ${\cal B}$ cycles:
here ${\cal A}$ can be chosen as a cut,
and ${\cal B}$ is represented as a thick dash-dot line.
Each cross {\bf x} (one in each sheet) represents a point $p\in{\cal E}$,
such that $z_k(p)=z_k$.}
\end{figure}

We are going to describe briefly the sheet geometry,
see fig.(\ref{sheetszk}) for a better vizualization.

The equation $z_k(p)=z_k$ has $r_k+s_k$ solutions for $p$,
i.e. the curve ${\cal E}$ is divided into $r_k+s_k$ sheets.
When $z_k\to\infty$, $r_k$ solutions approach $p_{\infty+}$, while $s_k$
approach $p_{\infty-}$.
In other words, $r_k$ sheets contain $p_{\infty+}$, while $s_k$ sheets
contain $p_{\infty-}$.

Let us denote ${\cal C}_k$ the contour which separates the reunion of sheets
containing $p_{\infty+}$ from the reunion of sheets containing $p_{\infty-}$,
oriented such that it turns around $p_{\infty+}$ in the positive direction.
That contour ${\cal C}_k$ will play an important role later.
Note that ${\cal C}_k$ is not necessarily connected
(as in our example fig. \ref{sheetszk}).

Let us also denote:\par
$\bullet$ $p_{+j,k}(z_k)$ ($j=1,\dots,r_k$) all the solutions of $z_k(p)=z_k$
which are on the same side of ${\cal C}_k$ than $p_{\infty+}$,\par
$\bullet$ $p_{-j,k}(z_k)$ ($j=1,\dots,s_k$) all the solutions of $z_k(p)=z_k$
which are on the same side of ${\cal C}_k$ than  $p_{\infty-}$.
\label{defpplusjminusj}

\subsection{Filling fractions}

It is well known that the discontinuities of $W_0(z_0)$, i.e.
the discontinuities of $Z_{-1}(z_0)$ along
the $Z_0$-cuts in the physical sheet are related to the large $N$ average
density
of eigenvalues of the matrix $M_0$:
\beq
\rho_0(z_0) = -{1\over 2i\pi } (W_0(z_0+i0)-W_0(z_0-i0))
\eeq
The support of $\rho_0$ is the set of cuts of $Z_{-1}(z_0)$, with endpoints
the $z_0(e_{0,i})$'s which belong to the physical sheet.

The ratio of the average number of eigenvalues in a given connected component
of the support of $\rho_0$ to the total number $N$ of eigenvalues
is a contour integral:
\beq
{1\over T}\epsilon_{i}=\int_{z_0(e_{0,i})}^{z_0(e_{0,i+1})} \rho(z_0)dz_0
= - {1\over 2i\pi T} \oint z_{-1} dz_0
\eeq
where the contour of integration on ${\cal E}$ is one of the $z_0$--cuts
defined in section \ref{sectioncuts}, oriented in the clockwise direction.
$\epsilon_i$ is non-zero only if the contour is a non-trivial contour,
or if it encloses a pole $p_{\infty+}$ or  $p_{\infty-}$.

Therefore, there are at most $g+1$ possible values of $\epsilon_{i}$,
the support of the density has at most $g+1$ connected components.

It is possible to choose the potentials $V_1,\dots,V_\N$ such that all
the non trivial cycles ${\cal A}_i$ are in the physical sheet, and such
that the cycles ${\cal A}_i$ are all cuts.
The filling fractions in that case are the ${\cal A}$--cycles integrals:
\beq\label{deffillingAi}
\epsilon_{i}
:= -{1\over 2i\pi} \oint_{{\cal A}_i} z_{-1} dz_0
\eeq

For more generic potentials, we define the filling fractions
by \refeq{deffillingAi}, eventhough they don't really correspond to numbers
of eigenvalues.
The cuts are integer homological linear combination of ${\cal A}$--cycles.

We also define the chemical potentials as the ${\cal B}$--cycle integrals:
\beq\label{defGammai}
\Gamma_{i}
:= -\oint_{{\cal B}_i} z_{-1} dz_0
\eeq

\medskip

We will see in part \ref{otherresolvents} that this description remains valid
for the densities
of all matrices $M_k$ (and not only $M_0$).
We will prove (see \refeq{Wk}) that $\rho_k(z_k)$ the density of eigenvalues
of $M_k$ is the
discontinuity of $z_{k+1}$ (or equivalently $z_{k-1}$) along the cuts.

It is easy to prove, by integration by parts, that:
\beq
\oint_{{\cal A}_i} z_{-1} dz_0 = \oint_{{\cal A}_i} z_{k-1} dz_k
= - \oint_{{\cal A}_i} z_{k+1} dz_{k}
\eeq
\beq
\oint_{{\cal B}_i} z_{-1} dz_0 = \oint_{{\cal B}_i} z_{k-1} dz_k
= - \oint_{{\cal B}_i} z_{k+1} dz_{k}
\eeq
If we anticipate on section \ref{otherresolvents},
this proves that the filling fractions
are the same for all densities,
up to integer linear combinations.
If the support of the density of eigenvalues of $M_k$ has $m_k\leq g+1$
connected components $[a_{k,i},b_{k,i}]$ ($i=1,\dots,m_k$),
the filling fractions in each connected component are:
\beq
\int_{a_{k,i}}^{b_{k,i}} \rho_k(z_k) dz_k = {1\over T}\sum_{j=1}^{g+1}
A_{k,i,j}\, \epsilon_j
\eeq
where the coefficients $A_{k,i,j}$ are integers (possibly nul or negative).

\subsection{${\cal A}$ or ${\cal B}$ cycles fixed ?}
\label{cuthypothesis}

As we said before, the loop equation is not sufficient to determine all
the unknown coefficients of $E$.
Some additional hypothesis are needed,
two of them are often considered in the litterature:

\begin{itemize}

\item {\bf condition B}: if all the potentials are bounded from below,
and the partition function
is well defined, one is interested in finding large $N$ limits for the free
energy and various expectation values of traces of powers of the matrices.
In that approach, one has to find a solution of the loop equation which gives
an absolute minimum of the free energy.
The genus $g$ and the filling fractions $\epsilon_i$ are not known,
they are determined by the minimization condition $\d F/\d\epsilon_i =0$,
which reduces to:
\beq\label{conditionB}
\forall i=1,\dots,g \qquad \oint_{{\cal B}_i} z_1 dz_0 =0
\eeq
All the ${\cal B}$ cycle integrals must vanish.
That condition is sufficient to determine $g$ and all the $\epsilon_i$,
and it is sufficient to determine the unknown coefficients of $E$.

This allows to find the large $N$ limit of the free energy.
Subleading large $N$ corrections, i.e. the so-called topological large $N$
expansion of the free energy, exist only if $g=0$,
as was shown in \cite{BDE, eynm2m}.
If the genus $g$ is $\geq 1$, there exist asymptotics for the subleading
corrections to the free energy, which have no $1/N$ expansion,
they are oscillating functions.

\item {\bf condition A}: if one is only interested in the formal large $N$
expansion of the free energy, one has to find a solution of the loop equation
which corresponds to the perturbation of a given local extremum of the free
energy.
Therefore, $g$ and the $\epsilon_i$'s are fixed parameters which characterize
the minimum around which we perform the perturbative expansion (moduli).
The following equations
\beq\label{conditionA}
\forall i=1,\dots,g \qquad \oint_{{\cal A}_i} z_1 dz_0 = 2i\pi \epsilon_i
\eeq
are sufficient to determine all the unknown coefficients of $E$.
\end{itemize}

In the following, we will always assume that we have condition A,
unless specified.
Most of the results obtained for condition A, immediately translate
to condition B, by exchanging the roles of ${\cal A}$ and ${\cal B}$-cycles,
like the large $N$ free energy.
But some quantities, like the large $N$ 2--loop functions,
get contributions from the oscillating asymptotics,
and will not be computed in this article.
So, from now on, we assume condition A.

\subsection{Parametrization}

It is possible to parametrize the algebraic curve ${\cal E}$ in terms of
$\theta$ functions.

Given $g$ and a canonical basis
of non-trivial cycles ${\cal{A}}_i , {\cal{B}}_i (i=1,\dots,g)$,
it is known \cite{Farkas, Fay} that there exists a unique basis of
holomorphic 1-forms $d u_i$
on ${\cal E}$ such that:
\beq\label{defdu}
\oint_{{\cal A}_i} du_j = \delta_{ij}
\eeq
The matrix of ${\cal B}$ periods is defined as:
\beq\label{deftau}
\tau_{ij} := \oint_{{\cal B}_i} du_j = \tau_{ji}
\eeq
$\tau_{ij}$ is a symmetric $g\times g$ matrix, with positive imaginary part.

If we choose an arbitrary basepoint $p_0\in{\cal E}$, we define the Abel map:
\beq\label{defAbelmap}
\vec{u}(p) := \int_{p_0}^p d\vec{u}
\eeq
which defines an embedding of ${\cal E}$ into $\CC^g$.

For any $0\leq k \leq \N$, $z_k(p)$ is a function on ${\cal E}$,
with a pole of degree $r_k$
at $p_{\infty+}$, and a pole of degree $s_k$ at  $p_{\infty-}$.
$z_k$ must have $r_k+s_k$ zeroes, which we denote $p_{k,i}(0)$,
($i=1,\dots,r_k+s_k$).
The zeroes must satisfy:
\beq
\sum_{i=1}^{r_k+s_k} \vec{u}(p_{k,i}(0)) = r_k\vec{u}(p_{\infty+})
+ s_k\vec{u}(p_{\infty-})
\eeq

then:
\beq\label{Zkthetaallgenus}
z_k(p) = A_k {\prod_{i=1}^{r_k+s_k}
\theta(\vec{u}(p)-\vec{u}(p_{k,i}(0)) - \vec{z};\tau)
\over \theta(\vec{u}(p)-\vec{u}(p_{\infty+}) - \vec{z};\tau)^{r_k}
\theta(\vec{u}(p)-\vec{u}(p_{\infty-}) - \vec{z};\tau)^{s_k} }
\eeq
where $\theta$ is the Riemann theta function \cite{Farkas, Fay},
$\vec{z}$ is an arbitrary non-singular zero of $\theta$.
The ratio of the RHS and the LHS is a function on ${\cal E}$ with no pole,
therefore it must be a constant, which we call $A_k$.

We have a parametrization of ${\cal E}$ in terms of $\theta$ functions.

\subsection{Two-loop functions and Bergmann kernel}

\subsubsection{The Bergmann kernel}

The Bergmann kernel $B(p;p')$ is the unique bilinear differential
form on ${\cal E}\times{\cal E}$ with the following properties:
\begin{itemize}
\item $B(p;p')$, as a function of $p$, is a meromorphic form,
with only one double pole at $p=p'$ with no residue,
and such that in any local parameter $z(p)$ we have:
\beq
B(p;p') \mathop\sim_{p\to p'} {dz(p)dz(p')\over (z(p)-z(p'))^2}
\eeq
\item $\forall i=1,\dots,g\qquad\oint_{p\in{\cal A}_i} B(p,p') = 0$.
\end{itemize}

\subsubsection{The 2-point function}

For $0\leq k\leq \N$ and $0\leq l\leq \N$,
define the following meromorphic differential forms on ${\cal E}\times \CC$:
\beq\label{defBkl}
B_{k,l}(p;z) :=
\left.{\d z_{k-1}(p)\over \d V_l(z)}\right|_{z_k(p)} \,\,
dz_k(p)\, d z
\eeq
\beq\label{deftdBkl}
\td{B}_{k,l}(p;z) :=
\left.{\d z_{k+1}(p)\over \d V_l(z)}\right|_{z_k(p)} \,\,
dz_k(p)\, d z
\eeq

We have the ``thermodynamic identity'':
\beq\label{Bklthermo}
\td{B}_{k,l}(p;z) = - B_{k+1,l}(p;z)
\eeq
and $z_{k-1}+z_{k+1}=V'_k(z_k)$ implies:
\beq\label{BkltdBkl}
B_{k,l}(p;z)+\td{B}_{k,l}(p;z) =
-\delta_{k,l}{dz_l(p)\, d z\over (z-z_l(p))^2}
\eeq

We are going to prove below that:
\beq\label{BklBergmann}
\begin{array}{lr}
B_{k,l}(p;z) = -\sum_{j=1}^{s_l} B(p;p_{-j,l}(z)) & \quad 0\leq k\leq l \cr
B_{k,l}(p;z) = \sum_{j=1}^{r_l} B(p;p_{+j,l}(z)) & \quad l< k\leq \N \cr
\td{B}_{k,l}(p;z) = \sum_{j=1}^{s_l} B(p;p_{-j,l}(z)) & \quad 0\leq k<l \cr
\td{B}_{k,l}(p;z) = -\sum_{j=1}^{r_l} B(p;p_{+j,l}(z)) & \quad l\leq k\leq \N
\end{array}
\eeq
where $B$ is the Bergmann kernel, and the $p_{j,l}$'s have been defined \
in \ref{defpplusjminusj}.

\bigskip
{\noindent \bf Proof of \refeq{BklBergmann}}

First notice that \refeq{Bklthermo} and \refeq{BkltdBkl} imply:
\beq
\begin{array}{lr}
B_{k,l}(p;z) = B_{0,l}(p;z) & \quad 0\leq k\leq l \cr
B_{k,l}(p;z) = -\td{B}_{\N,l}(p;z) & \quad l< k\leq \N
\end{array}
\eeq
and:
\beq\label{dualB0BN}
B_{0,l}(p;z) + \td{B}_{\N,l}(p;z) =
-{dz_l(p)\, d z\over (z-z_l(p))^2}
\eeq

Consider the two-loop functions introduced in \refeq{defWkl2loop}:
\beq
W_{0;l}(z;z')
=\left< \tr{1\over z-M_0}\tr{1\over z'-M_l}\right>_{\rm conn}
= {B_{0,l}(p_{+1,0}(z);z')\over dz_0(p_{+1,0}(z)) dz'}
\eeq
\beq
W_{\N;l}(z;z')
=\left< \tr{1\over z-M_\N}\tr{1\over z'-M_l}\right>_{\rm conn}
= {\td{B}_{\N,l}(p_{-1,\N}(z);z')\over dz_\N(p_{-1,\N}(z)) dz'}
\eeq
The expectation values
$\left< \tr{1\over z-M_0}\tr{1\over z'-M_l}\right>$
and  $\left< \tr{1\over z-M_\N}\tr{1\over z'-M_l}\right>$
are well defined integrals for $z$ in the physical sheet,
therefore $B_{0,l}(p;z')$ can have no pole when $z'=z_l(p)$
if $p$ is in the $z_0$-physical sheet,
and $\td{B}_{\N,l}(p;z')$ can have no pole when $z'=z_l(p)$
if $p$ is in the $z_\N$-physical sheet.
Note that the derivatives $\d/\d V_l(z')$ are formally defined only
for large $z'$, which implies that:\par
$B_{0,l}(p;z')$ can have no pole when
$p=p_{+j,l}(z')$, $j=1,\dots,r_l$,\par
$\td{B}_{\N,l}(p;z')$ can have no pole when
$p=p_{-j,l}(z')$, $j=1,\dots,s_l$.\par
\noindent Since the RHS of \refeq{dualB0BN} has poles at all
$p=p_{\pm j,l}(z')$,
we must have:\par
$B_{0,l}(p;z')$ can have no pole when
$p=p_{+j,l}(z')$, $j=1,\dots,r_l$,\par
$B_{0,l}(p;z')$ has double poles when
$p=p_{-j,l}(z')$, $j=1,\dots,s_l$.\par
\noindent That implies that:
\beq
\overline{B}_l(p;z'):=B_{0,l}(p;z') + \sum_{j=1}^{s_l} B(p;p_{-j,l}(z'))
\eeq
has no pole when $z_l(p)=z'$.

It obeys the following properties:
\begin{itemize}

\item Since $W_0(z_0)$ behaves as $1/z_0$ near $\infty$ in the physical sheet,
$W_{0;l}(z_0;z')$ must behave as $O(1/z_0^2)$ near $p_{\infty+}$,
while $dz_0$ has a
double pole at $p_{\infty+}$. Therefore
$B_{0,l}(p;z')$ has no pole when $p\to p_{\infty+}$, and thus
$\overline{B}_l(p;z')$ has no pole when $p\to p_{\infty+}$.

\item Similarly, $\td{B}_{\N,l}(p;z')$ has no pole when $p\to p_{\infty-}$,
and with \refeq{dualB0BN}, that implies that
$B_{0,l}(p;z')$ has no pole when $p\to p_{\infty-}$, and thus
$\overline{B}_l(p;z')$ has no pole when $p\to p_{\infty-}$.

\item Near an endpoint $e_{k,i}$, $Z_{k-1}(z_k)$ has a square root singularity,
i.e. the derivative $\d z_{k-1}(p)/\d V_l(z')|_{z_k(p)}$ has an inverse
square root singularity,
i.e. a simple pole which is exactly compensated by the zero of $dz_k(p)$.
Therefore $B_{k,l}(p;z')$ has no pole when $p\to e_{k,i}$, and so for
$\overline{B}_l(p;z')$.

\item Near any other point, $z_{k-1}$ is an analytical function of $z_k$,
and thus $B_{k,l}$ is analytical.

\item $B_{k,l}$ must satisfy:
\beq
\oint_{p\in{\cal A}_i} B_{k,l}(p;z') = 0
\eeq
and so must $\overline{B}_l$.

\end{itemize}

Finaly we find that $\overline{B}_l$ is a meromorphic form with no pole,
with all its ${\cal A}$ cycle integrals vanishing, therefore:
$\overline{B}_l = 0$, QED.

\Remark

We have:
\beq
B(p,p') = d_p d_{p'} \ln{\theta(\vec{u}(p)-\vec{u}(p')-\vec{z})}
\eeq
Notice that
\beq\label{duidVB}
d u_i(p') =
{1\over 2i\pi}\oint_{p\in{\cal B}_i} B(p,p')
= dz_0(p') {1\over 2i\pi} {\d \Gamma_i \over \d V_0(z_0(p'))}
\eeq
Once $B$ is known,
\refeq{duidVB} give an explicit way of computing
the Abel map $\vec{u}(p)$.

\subsection{Abelian differential of the 3rd kind and temperature}

\subsubsection{Abelian differential of the 3rd kind}

There is a unique abelian differential $dS$ on ${\cal E}$ with only two simple
poles at $p_{\infty\pm}$, such that:
\beq\label{defdS}
\Res_{p_{\infty+}} dS = -  \Res_{p_{\infty-}} dS = 1
\virg
\forall i=1,\dots,g \quad \oint_{{\cal A}_i} dS=0
\eeq
It has the property that:
\beq\label{dSintB}
dS(p) = \int_{p'=p_{\infty-}}^{p_{\infty+}} B(p,p')
\eeq

We define:
\beq\label{defetai}
\eta_i:=\oint_{{\cal B}_i} dS
\eeq

And, given a basepoint $p_0$, we define the following multivalued
function on ${\cal E}$:
\beq\label{defLambda}
\Lambda(p) := \ee{-\int_{p_0}^p dS}
\eeq
$\Lambda$ has a simple pole at $p_{\infty+}$ and a single zero
at $p_{\infty-}$, therefore the following quantities are well defined:
\beq\label{defgamma}
\gamma:= \mathop{\rm lim}_{p\to p_{\infty+}} {z_0(p)\over \Lambda(p)}
\virg
\td\gamma:= \mathop{\rm lim}_{p\to p_{\infty-}} {z_\N(p)\Lambda(p)}
\eeq
Notice that the product $\gamma\td\gamma$ is independent of
the choice of the basepoint$p_0$.

\subsubsection{Derivatives with respect to $T$}

Consider the abelian differentials:
\beq\label{defdSk}
dS_k(p) := \left.{\d z_{k-1}(p)\over \d T}\right|_{z_k(p)} dz_k(p)
\virg
d\td{S}_k(p) := \left.{\d z_{k+1}(p)\over \d T}\right|_{z_k(p)} dz_k(p)
\eeq
We clearly have (from $z_{k+1}+z_{k-1}=V'_k(z_k)$ and thermodynamic identity):
\beq
dS_k(p)=-d\td{S}_k(p)=dS_{k+1}(p)
\eeq
Therefore $dS_k$ and $d\td{S}_k$ are independent of $k$.

They have the following properties:
\begin{itemize}
\item near $p_{\infty+}$, $z_{-1}\sim T/z_0 + O(z_0^{-2})$,
thus $dS_0\sim dz_0/z_0$,
i.e. $dS_0$ has a single pole at  $p_{\infty+}$, with residue $-1$.
\item Similarly $d\td{S}_\N$ has a single pole at  $p_{\infty-}$, with residue $-1$,
i.e. $dS_0$ has a single pole at  $p_{\infty-}$, with residue $+1$.
\item near an endpoint $e_{k,i}$, $Z_{k-1}(z_k)$ has a square root singularity,
thus $\left.{\d z_{k-1}\over \d T}\right|_{z_k}$ has an inverse square root
singularity,
i.e. a single pole at $e_{k,i}$, which is compensated by the single zero
of $dz_K$, and
therefore $dS_0$ has no pole at $e_{k,i}$.
\item near any other point, $z_{k}$ is a local parameter and $z_{k+1}$
is analytical in
$z_k$, i.e. $dS_0$ has no pole.
\item The ${\cal A}_i$ cycle integrals of $dS_0$ vanish.
\end{itemize}
There is a unique abelian differential with such properties,
it is the abelian differential of the third kind defined in \refeq{defdS}:
\beq\label{dSkdS}
d\td{S}_k(p) = - dS_k(p) = dS(p)
\eeq
In particular, we have that:
\beq
{\d \over \d T} \Gamma_i = \eta_i
\virg
{\d \over \d T} \epsilon_i = 0
\eeq

\subsection{Derivatives with respect to $\epsilon_i$ (condition A)}

Consider:
\beq\label{defduk}
du_{k,i}(p)
:= {1\over 2i\pi}\left.{\d z_{k-1}(p)\over \d \epsilon_i}\right|_{z_k(p)}
dz_k(p)
\virg
d\td{u}_{k,i}(p)
:= {1\over 2i\pi}\left.{\d z_{k+1}(p)\over \d \epsilon_i}\right|_{z_k(p)}
dz_k(p)
\eeq
We clearly have (from $z_{k+1}+z_{k-1}=V'_k(z_k)$ and thermodynamic identity):
\beq
du_{k,i}(p)=-d\td{u}_{k,i}(p)=du_{k+1,i}(p)
\eeq
Therefore $du_{k,i}$ and $d\td{u}_{k,i}$ are independent of $k$.

Following the same lines as in the previous section, we show that
$du_{k,i}$ and $d\td{u}_{k,i}$ have no poles, i.e. they are holomorphic
one-forms.
Moreover we have:
\beq
\oint_{{\cal A}_j} d\td{u}_{k,i} = {\d \epsilon_j\over \d\epsilon_i}
= \delta_{i,j}
\eeq
There exists a unique set of holomorphic one-forms with those properties,
it is the holomorphic forms $du_i$ introduced in \refeq{defdu},
therefore:
\beq\label{dukdu}
d\td{u}_{k,i} = -du_{k,i} = du_i
\eeq
This shows that:
\beq\label{dGammadepsilon}
{\d \Gamma_j\over \d\epsilon_i} = 2i\pi\, \tau_{i,j}
\eeq
where $\tau_{i,j}$ is the matrix of periods introduced in \refeq{deftau}.

\section{Large $N$ free energy}

\subsection{The large $N$ free energy}
The free energy $F$ defined in \refeq{defZF} has a large $N$ limit:
\beq
F=F^{(0)}+ O(N^{-2})
\eeq
We prove below, a generalization of the formula of
\cite{Bertolafreeenergy}:
\beq\label{leadingF}
\encadremath{
\begin{array}{lcl}
2F^{(0)} &  = &
\displaystyle {\sum_{k=0}^{\N}
\Res_{p_{\infty_+}} (V_k(z_k)-{1\over 2} z_k V'_k(z_k)) z_{k+1} dz_k
   } \cr
   && + T \mu +\sum_i \epsilon_i \Gamma_i-T^2( 1+\N \ln{T}) \cr
& = & \displaystyle {\sum_{k=0}^{\N}
\Res_{p_{\infty_-}} (V_k(z_k)-{1\over 2} z_k V'_k(z_k)) z_{k-1} dz_k
   } \cr
   && + T \mu +\sum_i \epsilon_i \Gamma_i-T^2( 1+\N \ln{T}) \cr
\end{array}
}\eeq
where $\mu$ is the generalized version of \cite{Bertolafreeenergy}
and is defined as follows for any $p\in{\cal E}$:
\bea\label{defmu}
\mu & := & \int_{p_{\infty+}}^{p} ({T\over z_0}-z_{-1})dz_0
 + \int_{p_{\infty-}}^{p} ({T\over z_\N}-z_{\N+1})dz_\N
 - T\ln{z_0(p)} - T\ln{z_\N(p)} \cr
&&  + \sum_{k=0}^{\N} V_k(z_k(p)) - \sum_{k=1}^{\N} z_{k-1}(p)z_k(p)
\eea
$\mu$ is independent of $p$ (indeed $d\mu$ is a telescopic sum which
cancels completely).

\medskip
{\noindent \bf Proof of \refeq{leadingF}:}
We remind that we assume condition A.
We define:
\bea\label{defK}
4K & := &  2T\mu +2\sum_i \epsilon_i \Gamma_i -4F^{(0)}
+ \sum_{k=0}^{\N} \Res_{p_{\infty+}} (V_k(z_k)-{1\over 2}
z_k z_{k+1}) z_{k+1} dz_k \cr
&& + \sum_{k=0}^{\N} \Res_{p_{\infty-}} (V_k(z_k)-{1\over 2}
z_k z_{k-1}) z_{k-1} dz_k
\eea
Notice that
$\Res_{p_{\infty+}} z_k z_{k-1} z_{k+1} dz_k
+\Res_{p_{\infty-}} z_k z_{k-1} z_{k+1} dz_k=0$, so that the expression in
\refeq{defK} is the same as in \refeq{leadingF}.

\medskip
Let us compute:
\bea
4{\d K\over \d V_l(z)}
& = & \sum_k \Res_{p_{\infty+}} {\d V_k(z_k)\over \d V_l(z)} z_{k+1} dz_k
+ \sum_k \Res_{p_{\infty-}} {\d V_k(z_k)\over \d V_l(z)} z_{k-1} dz_k
+ 4 T W_l(z) \cr
&   & + \sum_k \Res_{p_{\infty+}}
\left.{\d z_{k+1}\over \d V_l(z)}\right|_{z_k}
(V_k(z_k)-z_k z_{k+1}) dz_k \cr
&   & + \sum_k \Res_{p_{\infty-}}
\left.{\d z_{k-1}\over \d V_l(z)}\right|_{z_k}
(V_k(z_k)-z_k z_{k-1}) dz_k \cr
&   & + 2T{\d \mu\over \d V_l(z)}
+2\sum_i \epsilon_i {\d \Gamma_i \over \d V_l(z)}
\eea
we introduce the following multivalued functions:
\beq
\zeta_k(p) :=\int_{p_{\infty+}}^p
\left.{\d z_{k+1}\over \d V_l(z)}\right|_{z_k} dz_k
\virg
\td\zeta_k(p) :=\int_{p_{\infty-}}^p
\left.{\d z_{k-1}\over \d V_l(z)}\right|_{z_k} dz_k
\eeq
By integration by parts, and using \refeq{dVdV}, we have:
\bea
4{\d K\over \d V_l(z)}
& = & -\Res_{p_{\infty+}} {z_{l+1}\over z-z_l} dz_l
- \Res_{p_{\infty-}} {z_{l-1}\over z-z_l} dz_l  + 4 TW_l(z) \cr
&   & - \sum_k \Res_{p_{\infty+}} \zeta_k
((V'_k(z_k)- z_{k+1})dz_k - z_k dz_{k+1}) \cr
&   & - \sum_k \Res_{p_{\infty-}} \td\zeta_k
((V'_k(z_k)- z_{k-1})dz_k - z_k dz_{k-1}) \cr
&   & + 2T{\d \mu\over \d V_l(z)}
+2\sum_i \epsilon_i {\d \Gamma_i \over \d V_l(z)} \cr
& = & -\Res_{p_{\infty+}} {z_{l+1}\over z-z_l} dz_l
- \Res_{p_{\infty-}} {z_{l-1}\over z-z_l} dz_l  + 4 TW_l(z) \cr
&   & - \sum_k \Res_{p_{\infty+}} \zeta_k ( z_{k-1} dz_k - z_k dz_{k+1}) \cr
&   & - \sum_k \Res_{p_{\infty-}}
\td\zeta_k ( z_{k+1} dz_k - z_k dz_{k-1}) \cr
&   & + 2T{\d \mu\over \d V_l(z)}
+2\sum_i \epsilon_i {\d \Gamma_i \over \d V_l(z)} \cr
& = & -\Res_{p_{\infty+}} {z_{l+1}\over z-z_l} dz_l
- \Res_{p_{\infty-}} {z_{l-1}\over z-z_l} dz_l  + 4 TW_l(z) \cr
&   &
+ \Res_{p_{\infty+}} \zeta_\N z_{\N} dz_{\N+1}
-  \zeta_{-1} z_{-1} dz_{0}
- \sum_{k=0}^{\N} (\zeta_{k}-\zeta_{k-1}) z_{k-1} dz_{k}  \cr
&   &
+ \Res_{p_{\infty-}} \td\zeta_0 z_{0} dz_{-1}
- \td\zeta_{\N+1} z_{\N+1} dz_{\N}
- \sum_{k=0}^{\N} (\td\zeta_{k}-\td\zeta_{k+1}) z_{k+1} dz_{k}  \cr
&   & + 2T{\d \mu\over \d V_l(z)}
+2\sum_i \epsilon_i {\d \Gamma_i \over \d V_l(z)}
\eea
Near $p_{\infty+}$ we have $z_{-1} \sim T/z_0 + O(z_0^{-2})$, therefore
$z_0\sim T/z_{-1} +O(1)$, and thus $\zeta_{-1}$ has a zero at $p_{\infty+}$.
That implies that:
\beq
\Res_{p_{\infty+}} \zeta_{-1} z_{-1} dz_{0}  =0
\virg
\Res_{p_{\infty-}} \td\zeta_{\N+1} z_{\N+1} dz_{\N} =0
\eeq
Moreover, notice that
$\left.{\d z_{k}\over \d V_l(z)}\right|_{z_{k-1}} dz_{k-1}
=-\left.{\d z_{k-1}\over \d V_l(z)}\right|_{z_{k}} dz_{k}$, therefore:
\beq
\zeta_{k}(p)-\zeta_{k-1}(p) = \int_{p_{\infty+}}^p
\left(\left.{\d z_{k+1}\over \d V_l(z)}\right|_{z_k}
+\left.{\d z_{k-1}\over \d V_l(z)}\right|_{z_k}\right) dz_k
= \int_{p_{\infty+}}^p {\d V'_k(z_k)\over \d V_l(z)} dz_k
= { \delta_{k,l} \over z_l(p)-z}
\eeq
Thus:
\bea
4{\d K\over \d V_l(z)}
& = & \Res_{p_{\infty+}} {z_{l-1}-z_{l+1}\over z-z_l}  dz_l
+\Res_{p_{\infty-}} {z_{l+1}-z_{l-1}\over z-z_l} dz_l  + 4T W_l(z) \cr
&   & + \Res_{p_{\infty+}} \zeta_\N z_{\N} dz_{\N+1}
+ \Res_{p_{\infty-}} \td\zeta_0 z_0 dz_{-1}
 + 2T{\d \mu\over \d V_l(z)}
 + 2\sum_i \epsilon_i {\d \Gamma_i \over \d V_l(z)} \cr
& = & \Res_{p_{\infty+}} {z_{l-1}-z_{l+1}\over z-z_l}  dz_l
+\Res_{p_{\infty-}} {z_{l+1}-z_{l-1}\over z-z_l} dz_l  + 4 TW_l(z) \cr
&   & - \Res_{p_{\infty-}} \zeta_\N z_{\N} dz_{\N+1}
- \Res_{p_{\infty+}} \td\zeta_0 z_0 dz_{-1} + 2T{\d \mu\over \d V_l(z)}
+2\sum_i \epsilon_i {\d \Gamma_i \over \d V_l(z)}\cr
&   & + \sum_i {1\over 2i\pi}\oint_{{\cal A}_i}
{\rm disc}_{{\cal A}_i}\, (\td\zeta_0 z_0 dz_{-1})
+ \sum_i {1\over 2i\pi}\oint_{{\cal A}_i}
{\rm disc}_{{\cal A}_i}\, (\zeta_\N z_\N dz_{\N+1}) \cr
&   & - \sum_i {1\over 2i\pi}\oint_{{\cal B}_i}
{\rm disc}_{{\cal B}_i}\, (\td\zeta_0 z_0 dz_{-1})
- \sum_i {1\over 2i\pi}\oint_{{\cal B}_i}
{\rm disc}_{{\cal B}_i}\, (\zeta_\N z_\N dz_{\N+1}) \cr \cr
\eea
where we have used Riemann bilinear identity, and disc means taking discontinuity
of the considered multivalued function when crossing cycles.

\medskip

Now compute ${\d\mu/\d V_l(z)}$, all terms cancel but:
\beq
{\d \mu\over \d V_l(z)} = -{1\over z-z_l(p)}
- \int_{p_{\infty+}}^{p} \left.{\d z_{-1}\over \d V_l(z)}\right|_{z_0} dz_0
- \int_{p_{\infty-}}^{p} \left.{\d z_{\N+1}\over \d V_l(z)}\right|_{z_\N}
dz_\N
\eeq
which is independent of $p$.
In particular for $p=p_{\infty\pm}$, this proves that
$\td\zeta_0(p_{\infty+})$ and $\zeta_\N(p_{\infty-})$ are finite,
and:
\beq\label{dotmu}
{\d \mu\over \d V_l(z)}
=\td\zeta_0(p_{\infty+})=\zeta_\N(p_{\infty-})
\eeq
Thus:
\beq
\Res_{p_{\infty+}} \td\zeta_0 z_0 dz_{-1}
= T \td\zeta_0(p_{\infty+})
=  T{\d \mu\over \d V_l(z)}
= T\zeta_{\N}(p_{\infty-})
= \Res_{p_{\infty-}} \zeta_\N z_{\N} dz_{\N+1}
\eeq

\medskip

Now compute the discontinuities (crossing ${\cal B}_i$ is equivalent to going around 
${\cal A}_i$):
\beq
{\rm disc}_{{\cal B}_i}\, \td\zeta_0
=\oint_{{\cal A}_i} \left.{\d z_{-1}\over \d V_l(z)}\right|_{z_0} dz_0
=-2i\pi{\d \epsilon_i \over \d V_l(z)}
=0
\eeq
this implies that $\td\zeta_0$ has no discontinuity along ${\cal B}_i$, and
\beq
{\rm disc}_{{\cal A}_i}\, \td\zeta_0
= \oint_{{\cal B}_i} \left.{\d z_{-1}\over \d V_l(z)}\right|_{z_0} dz_0
= -{\d \Gamma_i \over \d V_l(z)}
\eeq
this implies that $\td\zeta_0$ has a constant discontinuity along ${\cal A}_i$.
Moreover $z_0$ and $dz_{-1}$ are monovalued, i.e. they have no discontinuity along 
${\cal A}_i$ or ${\cal B}_i$.
Thus:
\beq
\oint_{{\cal A}_i} {\rm disc}_{{\cal A}_i}\, (\td\zeta_0 z_0 dz_{-1})
= -{\d \Gamma_i \over \d V_l(z)} \oint_{{\cal A}_i} z_0 dz_{-1}
= -2i\pi \epsilon_i {\d \Gamma_i \over \d V_l(z)}
\eeq
Using similar arugments for $z_\N$, we arrive at:
\beq\label{dFdVl}
4{\d K\over \d V_l(z)}
 =
\Res_{p_{\infty+}} {z_{l-1}-z_{l+1}\over z-z_l}  dz_l
+\Res_{p_{\infty-}} {z_{l+1}-z_{l-1}\over z-z_l} dz_l  + 4T W_l(z)
\eeq
\medskip
When $l=0$, by definition \refeq{defZk}, we have
\bea
T W_0(z)
& = & \Res_{p_{\infty+}} {1\over z-z_0}z_1 dz_0
= -\Res_{p_{\infty+}} {1\over z-z_0}z_{-1} dz_0 \cr
& = & \Res_{p_{\infty-}} {1\over z-z_0}z_{-1} dz_0
= -\Res_{p_{\infty-}} {1\over z-z_0}z_{1} dz_0
\eea
therefore:
\beq
{\d K\over \d V_0(z)} = 0
\eeq
which proves that $K$ is independent of $V_0$.
In particular we can choose $V_0$ quadratic, and then we integrate $M_0$ out,
i.e. we reduce the problem of a chain of lenght $\N$ with potentials $V_0,\dots,V_\N$
to a chain of lenght $\N-1$ with potentials $V_1-{z_1^2\over 2},\dots,V_\N$.
It is easy to check directly that:
\beq
F_{\N}({z_0^2\over 2},V_1,\dots,V_\N) 
= F_{\N-1}(V_1-{z_1^2\over 2},\dots,V_\N) - {T^2\over 2}\ln{T}
\eeq
and thus:
\beq
K_\N(V_0,\dots,V_\N) = K_{\N}({z_0^2\over 2},V_1,\dots,V_\N) 
= K_{\N-1}(V_1-{z_1^2\over 2},\dots,V_\N)  + {T^2\over 2}\ln{T}
\eeq
Therefore, by recursion on $\N$, we find that $K$ is independent of all $V's$.
$K$ could still depend on $T$ and the $\epsilon_i$'s.
$K$ has been computed for $\N=1$ and $\N=0$ \cite{Bertolafreeenergy}, and we find:
\beq
K= {T^2\over 2}( 1+\N \ln{T})
\eeq
which is the same result as if all potentials are chosen gaussian 
(see section \ref{gaussian}).

QED

\Remark
En route, we have proved that for all $l$, the resolvent of the
$l^{\rm th}$ matrix is:
\bea\label{WldFdVl}
W_l(z)
& = &  {1\over T}\Res_{p_{\infty+}} {1\over z-z_l} z_{l+1} dz_l
 =   -{1\over T}\Res_{p_{\infty+}} {1\over z-z_l} z_{l-1} dz_l  \cr
& = & {1\over T}\Res_{p_{\infty-}} {1\over z-z_l} z_{l-1} dz_l
= -{1\over T}\Res_{p_{\infty-}} {1\over z-z_l} z_{l+1} dz_l
\eea

\bigskip

\Remark
we have:
\beq\label{alternatemu}
\mu = -T\ln{\gamma\td\gamma} + \sum_i \epsilon_i \eta_i
+ \Res_{p_{\infty+}} \left( \sum_{k=0}^{\N} V_k(z_k)
- \sum_{k=1}^{\N} z_{k-1}z_k  \right)dS
\eeq
where $dS$ and $\gamma$ and $\td\gamma$ and $\eta_i$
are defined in \refeq{defdS} and \refeq{defgamma}.

\medskip

{\noindent \bf Proof of \refeq{alternatemu}:}
Consider the functions:
\beq\label{defphi0}
\phi_0(p)  :=  \int_{p_{\infty+}}^{p} ({T\over z_0}-z_{-1})dz_0
\virg
\phi_\N(p)  :=  \int_{p_{\infty-}}^{p} ({T\over z_\N}-z_{\N+1})dz_\N
\eeq
They satisfy:
\beq
\Res_{p_{\infty+}} \phi_0 dS = 0
\virg
\Res_{p_{\infty-}} \phi_\N dS = 0
\eeq
and they have the following discontinuities along ${\cal A}_i$
or ${\cal B}_i$:
\beq
{\rm disc}_{{\cal A}_i}\, \phi_0 = -{\rm disc}_{{\cal A}_i}\, \phi_\N
= \Gamma_i
\virg
{\rm disc}_{{\cal B}_i}\, \phi_0 = -{\rm disc}_{{\cal B}_i}\, \phi_\N
= 2i\pi\, \epsilon_i
\eeq

From \refeq{defmu} and \refeq{defdS}, we have:
\bea
\mu & = & \Res_{p_{\infty+}} \mu dS
 =  \Res_{p_{\infty+}} \phi_0 dS
 + \Res_{p_{\infty+}} \phi_\N dS   - T\ln{\gamma\td\gamma} \cr
&&  + \Res_{p_{\infty+}} \left( \sum_{k=0}^{\N} V_k(z_k(p))
- \sum_{k=1}^{\N} z_{k-1}(p)z_k(p) \right) dS
\eea
We need to compute $\Res_{p_{\infty+}} \phi_\N dS$, we use Riemann's
bilinear identity:
\bea
\Res_{p_{\infty+}} \phi_\N dS & = & -\Res_{p_{\infty-}} \phi_\N dS
    + \sum_i {1\over 2i\pi}\oint_{{\cal A}_i} {\rm disc}_{{\cal A}_i}\,
    \phi_\N dS
    - \sum_i {1\over 2i\pi}\oint_{{\cal B}_i} {\rm disc}_{{\cal B}_i}\,
    \phi_\N dS \cr
& = & \sum_i \eta_i \epsilon_i
\eea
QED.

\subsection{Derivatives with respect to $\epsilon_i$}

We have:
\beq\label{dmudepsilon}
{d \mu\over d\epsilon_i} = \eta_i
\eeq

indeed, from \refeq{defmu}, we have:
\beq
{\d\mu\over \d \epsilon_i} =
-\int_{p_{\infty+}}^p \left.{\d z_{-1}\over \d \epsilon_i}\right|_{z_0} dz_0
-\int_{p_{\infty-}}^p \left.{\d z_{\N+1}\over \d \epsilon_i}\right|_{z_\N} dz_\N
\eeq
using \refeq{dukdu} we have:
\beq
{\d\mu\over \d \epsilon_i} =
2i\pi \int_{p_{\infty+}}^{p_{\infty-}} du_i = \eta_i
\eeq

\bigskip

We have:
\beq\label{dFdepsilon}
{d F^{(0)}\over d\epsilon_i} = \Gamma_i
\eeq
Indeed, from \refeq{leadingF}, and using \refeq{dukdu}, we have:
\bea
4{\d F^{(0)}\over \d \epsilon_i}
& = & 2T\eta_i + 2\Gamma_i + 2\sum_j \epsilon_j \tau_{i,j}
+ \sum_k \Res_{p_{\infty+}} (V_k(z_k)-z_k z_{k+1}) du_i \cr
&& - \sum_k \Res_{p_{\infty-}} (V_k(z_k)-z_k z_{k-1}) du_i \cr
\eea
Let us introduce the multivalued function:
\beq
u_i(p):=\int_{p_{\infty+}}^p du_i
\eeq
its discontinuities along the ${\cal A}$ and ${\cal B}$ cycles are:
\beq
{\rm disc}_{{\cal A}_j} u_i = \tau_{i,j}
\virg
{\rm disc}_{{\cal A}_j} u_i = \delta_{i,j}
\eeq
After an integration by parts, we have:
\bea
4{\d F\over \d \epsilon_i}
& = & 2T\eta_i + 2\Gamma_i + 2\sum_j \epsilon_j \tau_{i,j}
- \sum_k \Res_{p_{\infty+}} u_i(V'_k(z_k)dz_k-z_{k+1}dz_k - z_k dz_{k+1}) \cr
&& + \sum_k \Res_{p_{\infty-}} u_i (V'_k(z_k)dz_k-z_{k-1}dz_k-z_kdz_{k-1}) \cr
& = & 2T\eta_i + 2\Gamma_i + 2\sum_j \epsilon_j \tau_{i,j}
- \sum_k \Res_{p_{\infty+}} u_i(z_{k-1}dz_k - z_k dz_{k+1}) \cr
&& + \sum_k \Res_{p_{\infty-}} u_i (z_{k+1}dz_k-z_kdz_{k-1}) \cr
& = & 2T\eta_i + 2\Gamma_i + 2\sum_j \epsilon_j \tau_{i,j}
- \Res_{p_{\infty+}} u_i z_{-1}dz_0
+\Res_{p_{\infty+}} u_i z_\N dz_{\N+1} \cr
&& + \Res_{p_{\infty-}} u_i z_{\N+1}dz_\N
- \Res_{p_{\infty-}} u_i z_0dz_{-1} \cr
& = & 2T\eta_i + 2\Gamma_i + 2\sum_j \epsilon_j \tau_{i,j}
+\Res_{p_{\infty+}} u_i z_\N dz_{\N+1}
 + Tu_i({p_{\infty-}})
- \Res_{p_{\infty-}} u_i z_0dz_{-1} \cr
\eea
using Riemann's bilinear identity and \refeq{alternatemu},
we find \refeq{dFdepsilon}.

\subsection{Derivatives with respect to $T$}

We have:
\beq\label{dmudT}
{\d \mu\over \d T} = -\ln{\gamma\td\gamma}
\eeq

indeed, from \refeq{defmu}, we have:
\beq
{\d\mu\over \d T} =
\int_{p_{\infty+}}^p
({1\over z_0}-\left.{\d z_{-1}\over \d T}\right|_{z_0} ) dz_0
+\int_{p_{\infty-}}^p
({1\over z_\N}-\left.{\d z_{\N+1}\over \d T}\right|_{z_\N})dz_\N
-\ln{z_0} - \ln{z_\N}
\eeq
using \refeq{dSkdS} and \refeq{defLambda}, we have:
\bea
{\d\mu\over \d T} & = &
\int_{p_{\infty+}}^p ({dz_0\over z_0}-{d\Lambda\over \Lambda})
+\int_{p_{\infty-}}^p ({dz_\N\over z_\N}+{d\Lambda\over \Lambda})
-\ln{z_0} - \ln{z_\N} \cr
& = & \ln{z_0\over \Lambda} - \ln\gamma + \ln{z_\N\Lambda} -\ln{\td\gamma}
- \ln{z_0} - \ln{z_\N}
 =  -\ln{\gamma\td\gamma}
\eea

\bigskip

We have:
\beq\label{dFdT}
{d F^{(0)}\over dT} = \mu -T({\N+3\over 2}+\N\ln{T})
\eeq
Indeed, from \refeq{leadingF}, and using \refeq{dSkdS}, we have:
\bea
4{\d F\over \d T}
& = & 2\mu + 2T{\d\mu\over \d T} + 2\sum_i \epsilon_i \eta_i
+ \sum_k \Res_{p_{\infty+}} (V_k(z_k)-z_k z_{k+1}) dS \cr
&& - \sum_k \Res_{p_{\infty-}} (V_k(z_k)-z_k z_{k-1}) dS
- 4T(1+\N\ln{T}) - 2\N T
\eea
using \refeq{dmudT} and \refeq{alternatemu}, we find \refeq{dFdT}.

\Remark
The derivative of the free energy wrt to T  can be computed directely from
\refeq{probaweight}:
\beq
-{N^2}\left.{\d (F/T^2)\over \d T}\right|_{\epsilon_i/T} = {N\over T^2}
\left< \tr
\sum_{k=0}^\N V_k(M_k)
-\sum_{k=1}^\N M_k M_{k-1}
\right>
\eeq
The matrix integral is defined for fixed filling fractions,
i.e. for fixed $\epsilon_i/T$.
We can thus write:
\beq
T^2{\d (F/T^2)\over \d T}
+ \sum_i {\epsilon_i\over T} {{\d F\over \d \epsilon_i}}
= -{1\over N}
\left< \tr \sum_{k=0}^\N V_k(M_k) -\sum_{k=1}^\N M_k M_{k-1}
\right>
\eeq
Using the change of variable $\delta M_k=M_k$, we get the loop equation:
\beq\label{deltaM_klinear}
2T = {1\over N}\left<\tr M_k V'_k(M_k)- M_k (M_{k-1} +M_{k+1})  \right>
\eeq
which implies:
\beq
-{T^2}{\d (F/T^2)\over \d T}
-{1\over T}\sum_i \epsilon_i \Gamma_i= {1\over N}
\left< \tr
\sum_{k=0}^\N (V_k(M_k)- {1\over 2} M_k V'_k(M_k))\right> +(\N+1) T
\eeq
i.e.
\beq
2{F\over T}-{\d F\over \d T} -{1\over T}\sum_i \epsilon_i \Gamma_i=
\sum_{k=0}^\N \Res(V_k(M_k)- {1\over 2} M_k V'_k(M_k)) W_k dz_k + (\N+1) T
\eeq
which is equivalent to \refeq{dFdT}.\par
Note that \refeq{deltaM_klinear},
is nothing but the infinitesimal version of the
rescaling $M_k\to \alpha M_k$.
In particular one can choose $\alpha=\sqrt{T}$,
and get directely from \refeq{probaweight}:
\beq\label{Ft1}
F(g_{k,j},\td{g}_{k,j},\epsilon_i,T) =
T^2 F(g_{k,j}T^{j/2-1},\td{g}_{k,j}T^{j/2-1},\epsilon_i/T,1)
-{\N+1\over 2}T^2\ln{T}
\eeq
Taking the derivative of \refeq{Ft1} with respect to $T$,
gives again \refeq{dFdT}.

\section{Other observables, leading order}

\subsection{Resolvents, leading order}
\label{otherresolvents}

So far, we have found that $W_0(z_0)$ and $W_\N(z_\N)$ obey
algebraic equations.
We have been able to determine the resolvents for the matrices
at the extremities of the chain,
but not for intermediate matrices.
We are going to determine $W_k(z_k)$ for $0\leq k\leq\N$.

We start from \refeq{WldFdVl}:
\beq\label{WldFdVlbis}
W_k(z) = -{1\over 2i\pi T}\oint_{{\cal C}_k} {z_{k-1}\over z-z_k}  dz_k
\eeq
where the integration contour is
${\cal C}_k$ defined in \ref{defpplusjminusj}.
Indeed, the residue is a contour integral around $p_{\infty+}$,
and since equation \refeq{WldFdVl} was derived formally for large $z$
(i.e. order by order in the large $z$ expansion),
we have to assume that the contour of integration encloses
all the $r_k$ solutions of $z_k(p)=z$.
Moreover, the residue is the sum of poles at the $p_{+j,k}(z_k)$'s and at
$p_{\infty+}$:
\bea
{1\over 2i\pi}\oint_{{\cal C}_k} {z_{k-1}\over z_k-z}  dz_k
& = &
\sum_{j=1}^{r_k} z_{k-1}(p_{+j,k}(z_k))
+ {1\over 2i\pi}\oint_{{\cal C}_k} z_{k-1}{dz_k\over z_k}   \cr
& = &
\sum_{j=1}^{r_k} z_{k-1}(p_{+j,k}(z_k))
+ {1\over 2i\pi}\oint_{{\cal C}_k} z_{k-1}
{V''(z_{k-1})dz_{k-1}-dz_{k-2}\over V'_{k-1}(z_{k-1})-z_{k-2}}   \cr
& = &
\sum_{j=1}^{r_k} z_{k-1}(p_{+j,k}(z_k))
+{1\over 2i\pi}\oint_{{\cal C}_k}
{V''(z_{k-1})z_{k-1}\over V'_{k-1}(z_{k-1})} dz_{k-1}  \cr
& = &
\sum_{j=1}^{r_k} z_{k-1}(p_{+j,k}(z_k))
+ {g_{k-1,d_{k-1}}\over g_{k-1,d_{k-1}+1}}
\eea

Therefore:
\beq\label{Wk}
T W_k(z)
= {g_{k-1,d_{k-1}}\over g_{k-1,d_{k-1}+1}} \,\,
+ \sum_{j=1}^{r_k} z_{k-1}(p_{+j,k}(z))
= {g_{k+1,d_{k+1}}\over g_{k+1,d_{k+1}+1}} \,\,
+ \sum_{j=1}^{s_k} z_{k+1}(p_{-j,k}(z))
\eeq

Notice that we have:
\beq
\sum_{j=-s_k}^{r_k} z_{k-1}(p_{j,k}(z)) = s_k V'_k(z)
+ {g_{k+1,d_{k+1}}\over g_{k+1,d_{k+1}+1}}
- {g_{k-1,d_{k-1}}\over g_{k-1,d_{k-1}+1}}
\eeq

\subsection{2-loop functions, leading order}

From \refeq{BklBergmann} and \refeq{Wk} We find ($l\leq k$):
\beq\label{twoloopbergmann}
\encadremath{
\left<\Tr{1\over z_k-M_k}\Tr{1\over z_l-M_l}\right>_{\rm c} dz_k dz_l =
-\sum_{i=1}^{r_k}\sum_{j=1}^{r_l} B(p_{+i,k}(z_k),p_{+j,l}(z_l))
}\eeq

\subsection{2-point one loop functions}

\subsubsection{$W_{k,k+1}$}

Define the following polynomial of two variables:
\bea\label{defQk}
Q_{k,k+1} (z_k,z_{k+1})
& = & \Res_{\infty} dz_{0}\dots dz_{k-1}dz_{k+2}\dots dz_{\N}
{E(z_0,\dots,z_k,z_{k+1},\dots,z_\N)
\over \prod_{j=0}^\N (V_j(z_j)-z_{j+1}-z_{j-1})} \cr
& = & E(\ul{Z}_0,\dots,\ul{Z}_{k-1},z_k,z_{k+1},\ul{Z}_{k+2},\dots,\ul{Z}_\N)\cr
& = & (-1)^{r_k}\,\,
\prod_{j=0}^{k-1} g_{j,d_j+1}^{r_{j}} 
\,\,\prod_{j=k+1}^\N g_{j,d_j+1}^{s_{j}}
\,{\prod_{0\neq j=-s_k}^{r_k}
(z_{k+1}-z_{k+1}(p_{j,k}(z_{k})))} \cr
& = & (-1)^{s_{k+1}}\,\,
\prod_{j=0}^k g_{j,d_j+1}^{r_{j}} 
\,\,\prod_{j=k+2}^\N g_{j,d_j+1}^{s_{j}} 
\,{\prod_{0\neq j=-s_{k+1}}^{r_{k+1}}
(z_k-z_{k}(p_{j,k+1}(z_{k+1})))} \cr
\eea
where the $\ul{Z}_j(z_k,z_{k+1})$ are defined by:
\beq
\left\{\begin{array}{l}
\ul{Z}_k(z_k,z_{k+1})=z_k
\virg
\ul{Z}_{k+1}(z_k,z_{k+1})=z_{k+1} \cr
\ul{Z}_{j+1}(z_k,z_{k+1})= V'_l(\ul{Z}_{j}(z_k,z_{k+1}))
-\ul{Z}_{j-1}(z_k,z_{k+1}) \qquad j>k+1 \cr
\ul{Z}_{j-1}(z_k,z_{k+1})= V'_l(\ul{Z}_{j}(z_k,z_{k+1}))
-\ul{Z}_{j+1}(z_k,z_{k+1}) \qquad j<k \cr
\end{array}\right.
\eeq

We conjecture (proved in appendix B for $k=0$):
\bea\label{mixednextcorrelators}
1-W_{k,k+1}(z_k,z_{k+1})
& = & {Q_{k,k+1}(z_k,z_{k+1})
/\prod_{j=0}^{k-1} 
g_{j,d_j+1}^{r_{j}}
\prod_{j=k+2}^{\N} 
g_{j,d_j+1}^{s_{j}}
\over
\prod_{j=1}^{r_{k}} (z_{k+1}(p_{+j,k}(z_k))-z_{k+1})
\prod_{j=1}^{s_{k+1}} (z_{k}(p_{-j,k+1}(z_{k+1}))-z_{k}) }
\cr
& = & g_{k,d_k+1}^{r_k}\,
{\prod_{j=1}^{r_{k+1}} (z_{k}-z_{k}(p_{+j,k+1}(z_{k+1})))
\over \prod_{j=1}^{r_{k}} (z_{k+1}(p_{+j,k}(z_k))-z_{k+1}) }
 \cr
& = & g_{k+1,d_{k+1}+1}^{s_{k+1}}\, {\prod_{j=1}^{s_{k}} (z_{k+1}-z_{k+1}
(p_{-j,k}(z_{k})))
\over \prod_{j=1}^{s_{k+1}} (z_{k}(p_{-j,k+1}(z_{k+1}))-z_{k}) }
\cr
\eea

\Remark
\beq\label{Qknext}
Q_{k,k+1}(z_k,z_{k+1}) = Q_{k-1,k}(V'_k(z_k)-z_{k+1},z_k) 
\eeq
and we may conjecture that the spectral curves of the differential systems
defined in the appendix of \cite{BEH} are:
\beq
\det{\left(z_{k+1}{\mathbf 1}_{r_k+s_k} - {\cal D}_{k}(z_k)\right)}
= (-1)^{r_k}\, {Q_{k,k+1}(z_k,z_{k+1})
\over \prod_{j=0}^{k-1} g_{j,d_j+1}^{r_{j}}
\prod_{j=k+1}^{\N} g_{j,d_j+1}^{s_{j}}}
\eeq
so that the property \refeq{Qknext} would be nothing but the duality 
discovered in
 \cite{BEH}.

\subsubsection{The function $U$ in the large $N$ limit}

Since the function $U$ appears in the RHS of the master loop equation,
it is important to be able to compute it in the large $N$ limit.

Define (notice that $U_1=U$):
\beq
U_k(z_0,z_k,\dots,z_\N)  :=  \mathop\Pol_{z_1,\dots,z_\N} \,\,
W(z_0,\dots,z_\N)\, f_{k,\N}(z_k,\dots,z_\N)\,\prod_{j=1}^{k-1} z_j
\eeq

We shall prove that:
\beq\label{UkHkP}
\begin{array}{rl}
U_k(z_0,z_k,\dots,z_\N) & =  H_{k,\N}(z_k,\dots,z_{\N}) W_0(z_0) \cr
& - \sum_{l=k}^\N  {P(Z_0,\dots,Z_{l-1},z_l,\dots,z_\N)
-P(Z_0,\dots,Z_{l},z_{l+1},\dots,z_\N)
\over z_l-Z_l} H_{k,l-1}(z_k,\dots,z_{l-1}) \cr
& =  V'_\N(z_\N)-z_{\N-1}   \cr
& + \sum_{l=0}^{\N-1} {E(Z_0,\dots,Z_{l-1},z_l,\dots,z_\N)
-E(Z_0,\dots,Z_{l},z_{l+1},\dots,z_\N)
\over z_l-Z_l}
H_{k,l-1}(z_k,\dots,z_{l-1})
 \cr
\end{array}
\eeq
where
\beq\label{defHkl}
H_{k,l}(z_k,\dots,z_l) := \Pol_{z_0,\dots,z_\N} \prod_{j=0}^\N
{1\over z_j-Z_j}\,
f_{k,l}(z_k,\dots,z_l)\,\prod_{j<k} z_j\, \prod_{j>l} z_j
\eeq
i.e. $H_{k,l}$ is a polynomial in $z_k,\dots,z_l$, and satisfies:
\beq
H_{k+1,k}=1
\virg
H_{k,l}=0 \,\,{\rm if}\,\, k>l+1
\virg
H_{k,l}={V'_k(z_k)-V'_k(Z_k)\over z_k-Z_k}H_{k+1,l} - H_{k+2,l}
\eeq

In particular for $k=1$ \refeq{UkHkP} reduces to:
\beq\label{UlargeN}
\begin{array}{rcl}
 U(z_0,\dots,z_{\N}) & = & V'_\N(z_\N)-z_{\N-1}   \cr
&& + \sum_{k=0}^{\N-1}{E(Z_0,\dots,Z_{k},z_{k+1},\dots,z_\N)
-E(Z_0,\dots,Z_{k+1},z_{k+2},\dots,z_\N)
\over z_{k+1}-Z_{k+1}}
H_{1,k}(z_1,\dots,z_{k}) \cr
& = &  H_{1,\N}(z_1,\dots,z_{\N}) W_0(z_0) \cr
&& - \sum_{k=0}^{\N-1} {P(Z_0,\dots,Z_{k},z_{k+1},\dots,z_\N)
-P(Z_0,\dots,Z_{k+1},z_{k+2},\dots,z_\N)
\over z_{k+1}-Z_{k+1}}
H_{1,k}(z_1,\dots,z_{k}) \cr
\end{array}
\eeq

{\noindent \bf Proof of \refeq{UkHkP}:}

Let us define:
\beq
A_k(z_0,z_k,\dots,z_\N)
:=  \mathop\Pol_{z_1,\dots,z_{k-1},z_{k+1},\dots,z_\N} \,\,
\prod_{j=1}^{k-1} z_j \,\, f_{k+1,\N}(z_{k+1},\dots,z_\N) W(z_0,\dots,z_\N)
\eeq
\bea
S_k(z_0,z_k,\dots,z_\N) & := &
\mathop\Pol_{z_0,\dots,z_{k-1},z_{k+1},\dots,z_\N}
f_{0,k-1}(z_0,\dots,z_{k-1})\cr
&& f_{k+1,\N}(z_{k+1},\dots,z_\N) W(z_0,\dots,z_\N)
\eea
\bea
T_k(z_0,z_k,\dots,z_\N) & := &
\mathop\Pol_{z_0,\dots,z_\N}
f_{0,k-1}(z_0,\dots,z_{k-1})V'_k(z_k)\cr
&& f_{k+1,\N}(z_{k+1},\dots,z_\N) W(z_0,\dots,z_\N)
\eea
From \refeq{deffkl} we have:
\beq
U_k = \mathop\Pol_{z_k} V'_k A_k - U_{k+2}
\eeq
and from \refeq{Wkrechypothesis}, we have:
\bea
(z_k-Z_k)A_k = U_{k+1} - S_k(Z_0,\dots,Z_{k-1},z_k,\dots,z_\N)
\eea
That implies:
\bea\label{UkTk}
U_k & = & {V'_k(z_k)-V'_k(Z_k)\over z_k-Z_k} U_{k+1} - U_{k+2} \cr
&& - {T_k(Z_0,\dots,Z_{k-1},z_k,\dots,z_\N)
-T_k(Z_0,\dots,Z_{k},z_{k+1},\dots,z_\N)
\over z_k-Z_k}
\eea

It is easy to prove (by recursion),
that $f_{0,\N} - f_{0,k-1} V'_k(z_k) f_{k+1,\N}$ is linear in $z_k$,
in other words we can write:
\beq
f_{0,\N} = f_{0,k-1} V'_k(z_k) f_{k+1,\N}  + z_k B_k
\eeq
where $B_k$ is independent of $z_k$, therefore:
\beq
T_k(z_0,\dots,z_k,\dots,z_\N)-P(z_0,\dots,z_k,\dots,z_\N)
= {\rm independent\, of\,} z_k
\eeq

and thus \refeq{UkTk} can be rewritten:
\bea
U_k & = & {V'_k(z_k)-V'_k(Z_k)\over z_k-Z_k} U_{k+1} - U_{k+2} \cr
&& - {P(Z_0,\dots,Z_{k-1},z_k,\dots,z_\N)
-P(Z_0,\dots,Z_{k},z_{k+1},\dots,z_\N)
\over z_k-Z_k}
\eea
From which, the initial conditions $U_{\N+1}=W_0$, $U_{\N+2}=0$
(easily derived) imply \refeq{UkHkP}.

\subsubsection{The extremities correlator $W_{0,\N}$}

From \refeq{UlargeN} and \refeq{WkrechypothesisN}, we derive:
\beq\label{W0N}
W_{0,\N}(z_0,z_\N)=\sum_{k=0}^{\N-1}
{E(Z_0,\dots,Z_{k},\td{Z}_{k+1},\dots,\td{Z}_\N)
\over (Z_k-\td{Z}_k)(Z_{k+1}-\td{Z}_{k+1})}
\eeq
where:
\beq
Z_k := z_k(p_{+1,0}(z_0))
\virg
\td{Z}_k := z_k(p_{-1,\N}(z_\N))
\eeq

\section{Subleading epxansion}

The aim of this section is to generalize the calculation of
\cite{eynm2m,eynm2mg1},
and compute the next to leading $1/N^2$ term in the topological expansion.
In this purpose, we expand the various observables in a $1/N^2$ power series:
\beq\label{Zexpand}
Z_k(z_0)= Z_k^{(0)}(z_0) + {1\over N^2} Z_k^{(1)}(z_0) + \dots
\eeq
\beq\label{loopeqexpand}
P(z_0,\dots,z_\N)= P^{(0)}(z_0,\dots,z_\N)
+ {1\over N^2} P^{(1)}(z_0,\dots,z_\N) + \dots
\eeq
and so on.
Then we expand \refeq{masterloopeq} to order $1/N^2$:
\beq
\sum_{k=1}^\N Z_k^{(1)}(z_0) E_k
-P^{(1)}(z_0,Z_1^{(0)},\dots,Z_\N^{(0)})
=
U_{;0}^{(0)}(z_0,Z_1^{(0)},\dots,Z_\N^{(0)};z_0)
\eeq
where:
\beq\label{defEk}
E_k:= \left.{\d E^{(0)}(z_0,\dots,z_\N)\over \d z_k}\right|_{z_j=Z_j^{(0)}}
\eeq
From the definition of the $Z_k$'s \refeq{defZk}, we easily find:
\beq\label{ZkZ1H}
Z_k^{(1)}(z_0)=\overline{H}_{1,k-1}\,Z_1^{(1)}(z_0)
\eeq
where (the $H_{k,l}$'s have been defined in \refeq{defHkl}):
\beq\label{defHbarkl}
\overline{H}_{1,k} := H_{1,k}(Z_1^{(0)},\dots,Z_{k}^{(0)})
\eeq
We have:
\beq
\overline{H}_{1,-1}=0
\virg
\overline{H}_{1,0}=1
\virg
\overline{H}_{1,k}=V''_k(Z_k^{(0)})\overline{H}_{1,k-1}- \overline{H}_{1,k-2}
\eeq
i.e.
\beq
\overline{H}_{1,k} = \left.{\d^k f_{1,k}(z_1,\dots,z_k)
\over \d z_1\dots\d z_k}\right|_{z_j=Z_j^{(0)}}
\eeq
Therefore \refeq{loopeqexpand} reads:
\beq\label{Z1P1U}
Z_1^{(1)}(z_0) = {P^{(1)}(z_0,Z_1^{(0)},\dots,Z_\N^{(0)})
+ U_{;0}^{(0)}(z_0,Z_1^{(0)},\dots,Z_\N^{(0)};z_0)
\over
\sum_{k=1}^\N \overline{H}_{1,k-1}\, E_k}
\eeq
Similarly to what was done in \cite{eynm2m,eynm2mg1},
we determine the polynomial $P^{(1)}$,
by the condition that $Z_1^{(1)}(z_0)$ has singularities only at the
endpoints $e_{0,i}$
in the $z_0$--sheet. That condition is sufficient to determine all the
unknown coefficients of $P^{(1)}$.
Now, let us compute $U_{;0}^{(0)}$.

\subsection{Computation of $U_{;0}(z_0,\dots,z_\N;x)$}

Notice that there is no explicit depedence on $V_0$ in \refeq{defU2loop},
therefore:
\beq
U_{;0}(z_0,\dots,z_\N;z) = {\d U(z_0,\dots,z_\N)\over \d V_0(z)}
\eeq
For $k\geq 1$, write $z_k=Z_k^{(0)}+\zeta_k$,
and Taylor expand \refeq{UlargeN} in the $\zeta$'s:
\bea
 U(z_0,\dots,z_{\N}) & = & V'_\N(z_\N)-z_{\N-1}    \cr
&& - \sum_{k=1}^{\N}
(E_k+{1\over 2}\zeta_{k} E_{k,k}+\sum_{i=k+1}^{\N}\zeta_i E_{k,i})\cr
&& \qquad (\overline{H}_{1,k-1}
+\sum_{i=1}^{k-1}\zeta_i \overline{H}_{1,k-1;i})
\eea
where $E_k$ was defined in \refeq{defEk}, and $E_{k,l}$ is:
\beq\label{defEkl}
E_{k,l}:= \left.{\d^2 E^{(0)}(z_0,\dots,z_\N)
\over \d z_k \d z_l}\right|_{z_j=Z_j^{(0)}}
\eeq
$\overline{H}_{1,k}$ was defined in \refeq{defHbarkl} and:
\beq\label{defHbarkli}
\overline{H}_{1,k;i} := \left.{\d H_{1,k}(z_1,\dots,z_k)
\over \d z_i}\right|_{z_j=Z_j^{(0)}}
\eeq
From \refeq{defHkl} and \refeq{defHbarkl} we derive:
\beq\label{K1kiVHH}
\overline{H}_{1,k;i}={1\over 2}V'''_i(Z_i^{(0)})\,
\overline{H}_{1,i-1}\,\overline{H}_{i+1,k}
\eeq

Then take the $\d/\d V_0(z)$ derivative, using ${\d\zeta_k/\d V_0}
= -{\d Z_k/\d V_0}$,
and take the $\zeta\to 0$ limit:
\bea\label{U;0dEkdV0}
U_{;0}^{(0)}(Z_0,Z_1^{(0)},\dots,Z_\N^{(0)};z)
& = & \sum_{k=1}^{\N}
\left({\d E_{k}\over \d V_0(z)}
- {\d Z_k^{(0)}\over \d V_0(z)}{E_{k,k}\over 2}
- \sum_{j>k} {\d Z_j^{(0)} \over \d V_0(z)} E_{k,j}\right)
\overline{H}_{1,k-1} \cr
&& +\sum_{k=1}^{\N} E_k \left({\d\overline{H}_{1,k-1}\over \d V_0(z)}
- \sum_{j=1}^{k-1} {\d Z_j^{(0)}\over \d V_0(z)}
\overline{H}_{1,k-1;j}\right)
\eea

From \refeq{defZk}, one easily derives:
\beq
{\d Z_{k+1}^{(0)}\over \d V_0(z)} =
{\d Z_k^{(0)}\over \d V_0(z)} V''_k(Z_k^{(0)})
-{\d Z_{k-1}^{(0)}\over \d V_0(z)}
 = \overline{H}_{1,k} {\d Z_1^{(0)}\over \d V_0(z)}
\eeq
and from \refeq{defHbarkl}
\beq\label{dHkldV0eq}
{\d\overline{H}_{1,k-1}\over \d V_0(z)}
=\sum_{j=1}^{k-1} {\d Z_j^{(0)}\over \d V_0(z)} V'''_j(Z_j^{(0)})
\overline{H}_{1,j-1}\overline{H}_{j+1,k-1}
= 2 \sum_{j=1}^{k-1} {\d Z_j^{(0)}\over \d V_0(z)} \overline{H}_{1,k-1;j}
\eeq

\subsection{Computation of ${\d E_{k}/ \d V_0(z)}$}

Let us define $\overline{Z}_j(z_0,z_1)$:
\beq\label{defZbark}
\overline{Z}_0(z_0,z_1):=z_0
\virg
\overline{Z}_1(z_0,z_1):=z_1
\virg
\overline{Z}_{j+1}(z_0,z_1)
= V'_j(\overline{Z}_j(z_0,z_1))
-\overline{Z}_{j-1}(z_0,z_1)
\eeq
We have:
\beq\label{dZbardz1Hbar}
\left.{\d \overline{Z}_j(z_0,z_1)\over \d z_1}\right|_{z_1=Z_1^{(0)}}
= \overline{H}_{1,j-1}
\virg
\left.{\d^2 \overline{Z}_{k}(z_0,z_1)\over \d z_1^2}\right|_{z_1=Z_1^{(0)}}
= \sum_{j=1}^{k-1} V'''_j(Z_j^{(0)})\, {\overline{H}}_{1,j-1}^2\,
\overline{H}_{j+1,k-1}
\eeq
Consider the polynomials defined in \refeq{defQk}:
\bea\label{defQbis}
Q_{0,1}(z_0,z_1)
& := & E^{(0)}(\overline{Z}_0(z_0,z_1),\dots,\overline{Z}_\N(z_0,z_1)) \cr
& = & C (z_1-Z_1^{(0)}(z_0))\prod_{j=1}^{s_0} (z_1-z_1(p_{-j,0}(z_0))))
\eea
(where $C=-\prod_{j=1}^\N g_{j,d_j+1}^{s_j}$),
and take its derivative (using \refeq{dZbardz1Hbar}):
\beq
\left.{\d Q_{0,1}(z_0,z_1)\over \d z_1}\right|_{z_1=Z_1^{(0)}}
= \sum_{j=1}^{\N} E_j\, \overline{H}_{1,j-1}
=  C \prod_{j=1}^{s_0} (Z_1^{(0)}-z_1(p_{-j,0}(z_0)))
\eeq
That implies:
\bea
&& \sum_{j=1}^{\N} {\d E_j\over \d V_0(z)}\, \overline{H}_{1,j-1}
 + \sum_{j=1}^{\N} E_j\, {\d \overline{H}_{1,j-1}\over \d V_0(z)} \cr
& = &
\sum_{l=1}^{s_0}
\left(
{\d Z_1^{(0)}\over \d V_0(z)}
-\left.{\d z_1(p_{-j,0}(z_{0}))\over \d V_0(z)}\right|_{z_0}
\right)
{ \sum_{j=1}^{\N} E_j\, \overline{H}_{1,j-1}\over Z_1^{(0)}-z_1(p_{-l,0}(z_0))}
\eea
Plugging that into \refeq{U;0dEkdV0}, we get:
\bea\label{U;0sansdEkdV0}
U_{;0}^{(0)}(Z_0,Z_1^{(0)},\dots,Z_\N^{(0)};z)
& = &
\sum_{l=1}^{s_0}
\left({\d Z_1^{(0)}\over \d V_0(z)}-\left.{\d z_1(p_{-j,0}(z_{0}))
\over \d V_0(z)}\right|_{z_0}\right)
{ \sum_{j=1}^{\N} E_j\, \overline{H}_{1,j-1}
\over Z_1^{(0)}-z_1(p_{-l,0}(z_{0}))} \cr
&& - \sum_{k=1}^{\N} {\d Z_k^{(0)}\over \d V_0(z)}{E_{k,k}\over 2}\,
\overline{H}_{1,k-1}
 - \sum_{k=1}^{\N}\sum_{j>k} {\d Z_j^{(0)} \over \d V_0(z)} E_{k,j}\,
 \overline{H}_{1,k-1} \cr
&& - \sum_{k=1}^{\N} \sum_{j=1}^{k-1} {\d Z_j^{(0)}\over \d V_0(z)} E_k\,
\overline{H}_{1,k-1;j} \cr
& = &
\sum_{l=1}^{s_0}
\left({\d Z_1^{(0)}\over \d V_0(z)}-\left.{\d z_1(p_{-j,0}(z_{0}))
\over \d V_0(z)}\right|_{z_0}\right)
{ \sum_{j=1}^{\N} E_j\, \overline{H}_{1,j-1}
\over Z_1^{(0)}-z_1(p_{-l,0}(z_{0}))} \cr
&& - {\d Z_1^{(0)}\over \d V_0(z)}\sum_{k=1}^{\N}
\left({E_{k,k}\over 2}\,  {\overline{H}}_{1,k-1}^2
+\sum_{j>k}  E_{k,j}\, \overline{H}_{1,k-1}\, \overline{H}_{1,j-1} \right)\cr
&& - {1\over 2}{\d Z_1^{(0)}\over \d V_0(z)}\sum_{k=1}^{\N} \sum_{j=1}^{k-1}
E_k\,
V'''_j(Z_j^{(0)})\,
{\overline{H}}_{1,j-1}^2\,\overline{H}_{j+1,k-1} \cr
& = & -
\sum_{l=1}^{s_0}
\left.{\d z_1(p_{-j,0}(z_{0})\over \d V_0(z))}\right|_{z_0}
{\left(\sum_{j=1}^{\N} E_j\, \overline{H}_{1,j-1}\right)
\over Z_1^{(0)}-z_1(p_{-l,0}(z_{0}))} \cr
&& + {\d Z_1^{(0)}\over \d V_0(z)}
\sum_{l=1}^{s_0} { \left(\sum_{j=1}^{\N} E_j\, \overline{H}_{1,j-1}\right)
\over Z_1^{(0)}-z_1(p_{-l,0}(z_{0}))} \cr
&& - {1\over 2}{\d Z_1^{(0)}\over \d V_0(z)}\sum_{k=1}^{\N}\sum_{j=1}^{\N}
 E_{k,j}\, \overline{H}_{1,k-1}\, \overline{H}_{1,j-1} \cr
&& - {1\over 2}{\d Z_1^{(0)}\over \d V_0(z)}\sum_{k=1}^{\N} \sum_{j=1}^{k-1}
E_k\,
V'''_j(Z_j^{(0)})\,
{\overline{H}}_{1,j-1}^2\,\overline{H}_{j+1,k-1}
\eea

Take the second derivative of \refeq{defQbis}:
\bea\label{dQdz12}
\left.{\d^2 Q_{0,1}(z_0,z_1)\over \d z_1^2}\right|_{z_1=Z_1^{(0)}}
& = &  2 \left(\sum_{j=1}^{\N} E_j\, \overline{H}_{1,j-1}\right)
\sum_{l=1}^{s_0} { 1\over Z_1^{(0)}-z_1(p_{-l,0}(z_{0}))}  \cr
& = & \sum_{k=1}^{\N}\sum_{j=1}^{\N} E_{k,j}\, \overline{H}_{1,j-1}\,
\overline{H}_{1,j-1}
+ \sum_{k=1}^{\N} E_k \left.{\d^2 \overline{Z}_{k}(z_0,z_1)
\over \d z_1^2}\right|_{z_1=Z_1}
\eea
plugging \refeq{dQdz12} and \refeq{dZbardz1Hbar} into \refeq{U;0sansdEkdV0},
we get:
\beq\label{U0eq}
\encadremath{
{U_{;0}^{(0)}(Z_0,Z_1^{(0)},\dots,Z_\N^{(0)};z)
\over \sum_{j=1}^{\N} E_j\, \overline{H}_{1,j-1}}
=  -\sum_{l=1}^{s_0}
{1\over Z_1^{(0)}-z_1(p_{-l,0}(z_{0}))} \,
\left.{\d z_1(p_{-j,0}(z_{0}))\over \d V_0(z)}\right|_{z_0}
}\eeq
which can be compared to \cite{eynm2mg1}.

\subsection{Next to leading order}

Using \refeq{U0eq} into \refeq{Z1P1U}, and using \refeq{BklBergmann},
we get the $1/N^2$ correction for the resolvent:
\bea\label{Z1P1U0eq}
Z_1^{(1)}(z_0)
& = & {P^{(1)}(z_0,Z_1^{(0)},\dots,Z_\N^{(0)})
\over \sum_{k=1}^\N \overline{H}_{1,k-1}\, E_k}
-\sum_{l=1}^{s_0}
{1\over Z_1^{(0)}-z_1(p_{-l,0}(z_{0}))} \,
\left.{\d z_1(p_{-j,0}(z_{0}))\over \d V_0(z_0)}\right|_{z_0} \cr
& = &  \sum_{l=1}^{s_0}
{1\over z_1(p_{1,0}(z_0))-z_1(p_{-l,0}(z_0))} \,
{B(p_{-l,0}(z_0),p_{1,0}(z_0))\over dz_0(p_{1,0}(z_0))dz_0(p_{-l,0}(z_0))} \cr
&& +{P^{(1)}(z_0,Z_1^{(0)},\dots,Z_\N^{(0)})
\over \sum_{k=1}^\N \overline{H}_{1,k-1}\, E_k}
\eea
and we remind that $P^{(1)}$ is completely determined
by the condition that $Z_1^{(1)}(z_0)$ has singularities only at the
endpoints $e_{0,i}$ in the $z_0$--sheet, and has vanishing ${\cal B}$--cylce
integrals.

From there, it should be possible to extend the calculation of
\cite{eynm2m, eynm2mg1} to the chain of matrices, and compute the $1/N^2$
term in the free energy for the chain of matrices.
This will be left for a later work.

\section{Examples}
\label{examples}

\subsection{Example: one-cut asumption (genus zero)}

Let us assume that the genus is $g=0$.

\smallskip

It was already discussed in \cite{eynm2m} that for multimatrix models, the
so called one-cut asumption should be replaced by a genus zero asumption.
Indeed, in that case,
the number of $z_0$--endpoints, i.e. the number of zeroes of
$dz_0$ is equal to the number of sheets, therefore there is exactly one cut
in the physical sheet.
${\cal E}$ is in one to one correspondance with the complex plane $\CC$, 
and it is well known \cite{Farkas, Fay} that the parametrization \refeq{Zkthetaallgenus}
is rational, and with $s={p-p_{\infty-}\over p-p_{\infty+}}$, it can be written:
\beq\label{zkparamrational}
z_k(s) = \sum_{i=-s_k}^{r_k} \alpha_{k,i}\,\, s^i \, .
\eeq
$z_k(s)$ has two poles: one pole of degree $r_k$ at $s=\infty$,
and one pole of
degree $s_k$ at $s=0$.
This parametrization is identical to the one found in \cite{eynardchain}
by the biorthogonal polynomial's method.

The set of equations:
\beq
\left\{
\displaystyle
\begin{array}{l}
\displaystyle
 V'_k(z_k(s))=z_{k+1}(s)+z_{k-1}(s) \cr
\displaystyle
 V'_0(z_0(s))-z_1(s)\mathop\sim_{s\to \infty} {T\over \alpha_{0,1}s} \cr
\displaystyle
 V'_{\N}(z_\N(s))-z_{\N-1}(s)\mathop\sim_{s\to 0} {Ts \over \alpha_{\N,-1}} \cr
\displaystyle
 \alpha_{0,1}=\alpha_{\N,-1}=\gamma=\td\gamma
\end{array}
\right.
\eeq
is sufficient to determine all the $\alpha_{k,i}$
(one is free to impose $\alpha_{0,1}=\alpha_{\N,-1}$ because $s$ can
be changed into any constant times $s$).

The Abelian differential of the third kind is:
\beq
dS = -{ds\over s}
\eeq
The Bergmann kernel is:
\beq
B(s,s') = {ds\, ds'\over (s-s')^2}
\eeq

\subsection{Example: gaussian case}
\label{gaussian}

Consider all potentials quadratic (i.e. $d_k=1$):
\beq
V_k(z_k):={g_k\over 2} z_k^2
\eeq
A direct computation of the matrix integral \refeq{defZF} gives 
the free energy:
\beq\label{Fgaussien}
F=F^{(0)} = {T^2\over 2}\ln{D_{0,\N}} - {\N+1\over 2} T^2\ln{T}
\eeq

Define:
\beq
D_{k,l} := \det\pmatrix{
g_k & 1 &   & \cr
1 & \ddots & \ddots  & \cr
 &   \ddots & \ddots & 1\cr
 &    & 1 & g_l
}
\virg
D_{k,k-1}:=1
\virg
D_{k,k-2}:=0
\eeq

The function $f_{k,l}$ of \refeq{deffkl} is:
\beq
f_{k,l} = z_k\dots z_l \, D_{k,l}
\eeq
thus the polynomial $P$ is a constant:
\beq
P(z_0,\dots,z_\N) = D_{0,\N}
\eeq
and the polynomial $E$ is:
\beq
E(z_0,\dots,z_\N) = (g_0 z_0-z_1)(g_\N z_\N-z_{\N-1})-TP
\eeq
The leading order loop equations are thus:
\beq
z_{k+1}+z_{k-1}=g_k z_k
\virg
z_{-1}z_{\N+1}=TP
\eeq

Note that \refeq{maxgenus} implies that the genus is necessarily $g=0$,
and thus there is a rational parametrization of the form
\refeq{zkparamrational}, with $r_k=s_k=1$, namely:
\beq
z_{k} = \sqrt{T\over D_{0,\N}} \left(D_{0,k-1}s + D_{k+1,\N}s^{-1} \right)
\eeq
We find:
\beq
\gamma=\td\gamma = \sqrt{T\over D_{0,\N}}
\eeq
and:
\beq
\mu=T+T\ln{D_{0,\N}}-T\ln{T} = T-T\ln{\gamma\td\gamma} 
\eeq
and it is easy to check that \refeq{leadingF} coincides with \refeq{Fgaussien}.

\subsection{Loop equations of the 2 matrix model}

Let us take $T=1$.
The 2-matrix model corresponds to $\N=1$ we have:
\beq
f_{0,1}(z_0,z_1) = V'_0(z_0)V'_1(z_1)-z_0z_1
\virg
f_{1,1}(z_1)=V'_1(z_1)
\eeq
i.e.
\beq
P(z_0,z_1) = {1\over N} \left< \tr {V'_0(z_0)-V'_0(M_0)
\over z_0-M_0}{V'_1(z_1)-V'_1(M_1)\over z_1-M_1}\right> -1
\eeq
\beq
U(z_0,z_1;z) = \left< \tr {1\over z_0-M_0}{V'_1(z_1)-V'_1(M_1)
\over z_1-M_1}\tr {1\over z-M_0}\right>_{\rm c}
\eeq
and the master loop equation reads:
\beq
(V'_0(Z_0)-Z_1)(V'_1(Z_1)-Z_0)
- P(Z_0,Z_1) =
{1\over N^2} U(Z_0,Z_1;Z_0)
\eeq
with $Z_1=V'_0(z_0)-W_0(z_0)$.
We recover the equation of \cite{eynard,eynardchain,eynm2m}.

\subsection{The 3 matrix model}

Let us take $T=1$.
The 3-matrix model corresponds to $\N=2$, i.e.:
\beq
f_{0,2}(z_0,z_1,z_2) = V'_0(z_0)V'_1(z_1)V'_2(z_2)
-z_0z_1V'_2(z_2)-z_2z_1V'_0(z_0)
\eeq
\beq
f_{1,2}(z_1,z_2) = V'_1(z_1)V'_2(z_2) - z_1 z_2
\eeq
i.e.
\bea
P(z_0,z_1,z_2) & = & {1\over 2N} \left< \tr {V'_0(z_0)-V'_0(M_0)
\over z_0-M_0}{V'_1(z_1)-V'_1(M_1)\over z_1-M_1}
{V'_2(z_2)-V'_2(M_2)\over z_2-M_2}\right> \cr
&& + {1\over 2N} \left< \tr {V'_2(z_2)-V'_2(M_2)
\over z_2-M_2}{V'_1(z_1)-V'_1(M_1)\over z_1-M_1}
{V'_0(z_0)-V'_0(M_0)\over z_0-M_0}\right> \cr
&& - {1\over N} \left< \tr {V'_0(z_0)-V'_0(M_0)
\over z_0-M_0}\right>
- {1\over N} \left< \tr {V'_2(z_2)-V'_2(M_2)\over z_2-M_2}\right> \cr
\eea
\bea
U(z_0,z_1,z_2;z) & = & {1\over 2}\left< \tr
{1\over z_0-M_0}{V'_1(z_1)-V'_1(M_1)\over z_1-M_1}
{V'_2(z_2)-V'_2(M_2)\over z_2-M_2}
\tr {1\over z-M_0}\right>_{\rm c} \cr
&& + {1\over 2}\left< \tr
{V'_2(z_2)-V'_2(M_2)\over z_2-M_2}{V'_1(z_1)-V'_1(M_1)\over z_1-M_1}
{1\over z_0-M_0}
\tr {1\over z-M_0}\right>_{\rm c} \cr
&& - \left< \tr {1\over z_0-M_0}\tr {1\over z-M_0}\right>_{\rm c} \cr
\eea
and the master loop equation is:
\beq
(V'_0(Z_0)-Z_1)(V'_2(Z_2)-Z_1)
- P(Z_0,Z_1,Z_2)
={1\over N^2} U(Z_0,Z_1,Z_2;Z_0)
\eeq
with $Z_0=z_0$, $Z_1(z_0)=V'_0(z_0)-W_0(z_0)$ and $Z_2(z_0)=V'_1(Z_1)-Z_0$.

\bigskip
{\noindent \bf Large $N$ one loop functions}

We define:
\beq
\td{Z}_1(z_2)=V'_2(z_2)-W_2(z_2)
\virg
\td{Z}_0(z_2)=V'_1(\td{Z}_1(z_2))-z_2
\eeq
\beq
W_{\hat{0},1}(z_0,z_1) = \mathop\Pol_{z_0} V'_0(z_0) W_{0,1}(z_0,z_1)
\eeq
\beq
W_{1,\hat{2}}(z_1,z_2) = \mathop\Pol_{z_2} V'_2(z_2) W_{1,2}(z_1,z_2)
\eeq
\beq
W_{\hat{0},1,\hat{2}}(z_0,z_1,z_2) = \mathop\Pol_{z_0,z_2} V'_0(z_0)V'_2(z_2)
W(z_0,z_1,z_2)
\eeq

\refeq{W0N} reads:
\beq
W_{0,2}(z_0,z_2) = {1\over Z_1-\td{Z}_1}
\left({E(z_0,\td{Z}_1,z_2) \over z_0-\td{Z}_0}
- {E(z_0,Z_1,z_2)\over z_2-Z_2}\right)
\eeq
\refeq{UkHkP} reads:
\beq
U_2(z_0,z_2)= \mathop\Pol_{z_2} V'_2(z_2)W_{0,2}(z_0,z_2)
= {E(z_0,Z_1,z_2)\over z_2-Z_2}
\eeq
\refeq{UlargeN} reads:
\bea
U(z_0,z_1,z_2) & = & P_{1,2}(z_0,z_1,z_2) \cr
& = & V'_2(z_2)- z_1 + {E(z_0,z_1,z_2) - E(z_0,Z_1,z_2)\over z_1-Z_1} \cr
&& + {V'_1(z_1)-V'_1(Z_1)\over z_1-Z_1 }  {E(z_0,Z_1,z_2)\over z_2-Z_2}
\eea
\refeq{loopeqWzero} implies:
\bea
P_{1,1} &= &
{V'_1(z_1)-V'_1(Z_1)\over z_1-Z_1}W_{0,2}(z_0,z_2)
-{P_{1,2}(z_0,z_1,z_2)-P_{1,2}(z_0,Z_1,z_2)\over z_1-Z_1} \cr
& = &  {(V'_1-z_0-z_2)\over (Z_1-\td{Z}_1)} \left(
{E(z_0,\td{Z}_1,z_2)\over (z_1-\td{Z}_1) (z_0-\td{Z}_0)}
-{E(z_0,Z_1,z_2)\over (z_1-Z_1)(z_2-Z_2)}  \right) \cr
&& +1 - {E(z_0,z_1,z_2) \over (z_1-Z_1)(z_1-\td{Z}_1)}
\eea

We have the relationships:
\beq
(z_1-\td{Z}_1)P_{0,0}(z_0,z_1,z_2) = {E(z_0,\td{Z}_1,z_2)\over z_0-\td{Z}_0}
- W_{\hat{0},1,\hat{2}}(z_0,z_1,z_2)
\eeq
\bea
(Z_1-\td{Z}_1)W(z_0,z_1,z_2) & = &
{E(z_0,\td{Z}_1,z_2)\over (z_1-\td{Z}_1)(z_0-\td{Z}_0)}
- {E(z_0,Z_1,z_2)\over (z_1-Z_1)(z_2-Z_2)} \cr
&& + {(Z_1-\td{Z}_1)W_{\hat{0},1,\hat{2}}(z_0,z_1,z_2)
\over (z_1-Z_1)(z_1-\td{Z}_1)}
\eea
But so far, we have not been able to compute $W_{\hat{0},1,\hat{2}}$.
We conjecture:
\bea
&& (z_0+z_2-V'_1(z_1))W_{\hat{0},1,\hat{2}}(z_0,z_1,z_2)
 =  E(z_0,z_1,z_2) \cr
&& -(V'_0(z_0)-z_1-W_{\hat{0},1}(z_0,z_1))
(V'_2(z_2)-z_1-W_{1,\hat{2}}(z_1,z_2))
\eea
indeed, both sides are polynomials in $z_0$ and $z_2$ with the same degree
and same large $z_0$, $z_1$ and $z_2$ behaviours.

\section{Conclusion}

We have written the loop equations in a very explicit way.
To leading order the loop equations become algebraic, and using that
fact, we have derived many observables, like the free energy and many
correlators.

We have computed some mixed correlators and we have given conjecture for
others.
So far, the complete one--loop function has not been computed, and we
don't even have a conjecture (except in the 3-matrix model).
The mixed correlators are important in the study of boundary operators
\cite{kostov}.

It would also be interesting to generalize \cite{eynm2m} to the chain of
matrices, and find the next to leading large $N$ corrections to the free
energy.

\bigskip
{\em Aknowledgements}
The author would like to thank V. Kazakov for encouraging him to translate
in english (and add new results to) old notes written in french,
and to thank M. Bertola for fruitfull discussions and careful reading.
When this work was in progress, M. Bertola found independent proofs for
some of the formulae in section 5, which will be published later.

\setcounter{section}{0}
\appendix{Some usefull formulas with residues and polynomial parts}

We make an abundant use of the following formulas:
\beq
1 = \mathop\Pol_z {z\over z-Z} = \mathop\Pol_z {z\over z-M}
= -\Res {d z\over z-Z}
= -\Res {z dz \over (z-Z)(z-M)}
\eeq

\beq
\mathop\Pol_z {V'(z)\over z-Z} = {V'(z)-V'(Z)\over z-Z}
\virg
\Res { V'(z) d z\over z-Z} = -V'(Z)
\eeq

\beq\label{formulaPolVzZzM}
\mathop\Pol_z {V'(z)\over (z-Z)(z-M)}
= {1\over z-Z} \left({V'(z)-V'(M)\over z-M}-{V'(Z)-V'(M)\over Z-M}\right)
\eeq

\beq
\Res  {V'(z)dz\over (z-Z)(z-M)}
= -{V'(Z)-V'(M)\over Z-M}
\eeq

\beq\label{formulaVzM}
{V'(M)\over z-M} = {V'(z)\over z-M} - {V'(z)-V'(M)\over z-M}
= {V'(z)\over z-M} - \mathop\Pol_z {V'(z)\over z-M}
\eeq

\beq\label{formulaMzM}
{M\over z-M} = {z\over z-M} - 1
= {z\over z-M} - \mathop\Pol_z {z\over z-M}
\eeq

\appendix{Determination of $W_{0,1}$}
\label{appendixW01}

Define the following polynomial of $z_0$:
\beq
W_{\hat{0},1}(z_0,z_1) = \mathop\Pol_{z_0} V'_0(z_0) W_{0,1}(z_0,z_1)
\eeq

\refeq{loopeqWzero} implies the following identity
(to large $N$ leading order):
\beq
(z_1-Z_1) (1-W_{0,1}(z_0,z_1)) = z_1- V'_0(z_0) + W_{\hat{0},1}(z_0,z_1)
\eeq
which can also be written:
\beq
(1-{Z_1\over z_1}) (1-W_{0,1}(z_0,z_1)) = 1-{V'_0(z_0)
- W_{\hat{0},1}(z_0,z_1)\over z_1}
\eeq
Notice that the RHS is a polynomial in $z_0$.

take the log:
\beq\label{W01log}
\ln{(1-{Z_1\over z_1})} = -  \ln{(1-W_{0,1}(z_0,z_1))}
+ \ln{(1-{V'_0(z_0) - W_{\hat{0},1}(z_0,z_1)\over z_1})}
\eeq
take the fractionary part ($\Frac f := f - \Pol f$), we find:

\beq
-\ln{(1-W_{0,1}(z_0,z_1))} = \mathop{\Frac}_{z_0}
\left(\ln{(1-{Z_1(z_0)\over z_1})}\right)
\eeq
i.e.
\beq
W_{0,1}(z_0,z_1)
= 1-\exp{\left( - \mathop{\Frac}_{z_0}
\left(\ln{(1-{Z_1(z_0)\over z_1})}\right)\right)}
\eeq
in other words:
\bea
-\ln{(1-W_{0,1}(z_0,z_1))}
& = & {1\over 2i\pi} \oint_{p\in{\cal C}_0} {dZ_0(p)\over z_0-Z_0(p)}
\ln{(1-{Z_1(p)\over z_1})}   \cr
& = & {1\over 2i\pi} \oint_{{\cal C}_1} {dZ_1(p)\over Z_1(p)-z_1}
\ln{(1-{Z_0(p)\over z_0})}  \cr
& = & \mathop{\Frac}_{z_1} \left(\ln{(1-{Z_0(z_1)\over z_0})}\right) \cr
& = & \ln{(1-{Z_0(z_1)\over z_0})} - \mathop{\Pol}_{z_1}
\left(\ln{(1-{Z_0(z_1)\over z_0})}\right)
\eea
where the second equality is obtained by integration by parts.

In other words we have that:
\beq
(1-W_{0,1}(z_0,z_1))(z_1-Z_1(z_0)) = {\rm Polynomial\, in\, }z_0
\eeq
\beq
(1-W_{0,1}(z_0,z_1))\prod_{j=1}^{s_1}(z_0-Z_0(p_{-j,1}(z_1)))
= {\rm Polynomial\, in\, }z_1
\eeq
that implies that:
\beq
1-W_{0,1}(z_0,z_1)
= {Q_{0,1}(z_0,z_1)\over (z_1-Z_1(z_0))
\prod_{j=1}^{s_1}(z_0-Z_0(p_{-j,1}(z_1)))}
\eeq
where $Q_{0,1}(z_0,z_1)$ is some polynomial in both variables, of degree
$r_1+s_1,r_0+s_0$, which vanishes each time there exists $p$ such
that $z_0=Z_0(p)$ and $z_1=Z_1(p)$.
That implies that:
\bea
Q_{0,1}(z_0,z_1) & \propto & \prod_{j=1}^{r_1} (z_0-Z_0(p_{+j,1}(z_1)))
\prod_{j=1}^{s_1} (z_0-Z_0(p_{-j,1}(z_1))) \cr
& \propto & \prod_{j=1}^{r_0} (z_1-Z_1(p_{+j,0}(z_0))) \prod_{j=1}^{s_0}
(z_1-Z_1(p_{-j,0}(z_0)))  \cr
\eea
i.e. $Q_{0,1}$ must be equal to (same degree, same zeroes):
\beq
Q_{0,1}(z_0,z_1) \propto \oint dz_{2}\dots dz_{\N}
{E(z_0,z_1,z_2,\dots,z_\N)\over \prod_{j=1}^{\N-1} (z_{j+1}+z_{j-1}-V_j(z_j))}
\eeq
We thus have:
\beq
\encadremath{
W_{0,1}(z_0,z_1)
= 1- {Q_{0,1}(z_0,z_1)
\over (z_1-Z_1(z_0)) \prod_{j=1}^{s_1}(z_0-Z_0(p_{-j,1}(z_1)))}
}\eeq
i.e.
\beq
\encadremath{
W_{0,1}(z_0,z_1)
= 1+ g_{0,d_0+1}\, {\prod_{j=1}^{r_1}(z_0-Z_0(p_{+j,1}(z_1)))
\over (z_1-Z_1(p_{+1,0}(z_0)))}
}\eeq


\end{document}